\documentclass[journal]{IEEEtran}
\IEEEoverridecommandlockouts
\usepackage{cite}
\usepackage{amsmath,amssymb,amsfonts}
\usepackage{algorithmic}
\usepackage[ruled,vlined]{algorithm2e}
\usepackage{graphicx}
\usepackage{subcaption}
\usepackage{textcomp}
\usepackage{diagbox}
\usepackage{makecell}
\usepackage{threeparttable}
\usepackage{xcolor}
\allowdisplaybreaks
\usepackage[left=0.53in,right=0.53in,top=0.53in,bottom=0.53in]{geometry}
\usepackage[font=footnotesize,labelfont=footnotesize]{caption}
\def\BibTeX{{\rm B\kern-.05em{\sc i\kern-.025em b}\kern-.08em
    T\kern-.1667em\lower.7ex\hbox{E}\kern-.125emX}}
\begin{document}

\title{Joint Trajectory and Resource Optimization for \\Aerial RIS-assisted Integrated TNT Networks}

\author{Vangara Saiprudhvi,~\IEEEmembership{Graduate Student Member,~IEEE},  Sandeep Singh,~\IEEEmembership{Graduate Student Member,~IEEE},  Keshav Singh,~\IEEEmembership{Senior Member,~IEEE},  Hariharan Subramaniyam,~\IEEEmembership{Senior Member,~IEEE}, and Chih-Peng Li,~\IEEEmembership{Fellow,~IEEE}

\thanks{V. Saiprudhvi is with the Institute of Communications Engineering, National Sun Yat-sen University, Kaohsiung 80424, Taiwan, and also with the Department of Communication Engineering, School of Electronics Engineering, Vellore Institute of Technology, Vellore, Tamilnadu, India (e-mail: saiprudhvi000@gmail.com).}

\thanks{Hariharan S. is with the Department of Communication Engineering, School of Electronics Engineering, Vellore Institute of Technology, Vellore, Tamilnadu, India (e-mail: shariharan@vit.ac.in).}

\thanks{S. Singh, K. Singh, and C.-P. Li are with the Institute of Communications Engineering, National Sun Yat-sen University, Kaohsiung 80424, Taiwan (e-mail: d133070004@student.nsysu.edu.tw, keshav.singh@mail.nsysu.edu.tw, cpli@mail.nsysu.edu.tw).}
\thanks{This work has been submitted to IEEE Transactions on Wireless Communications for possible publication.}
\vspace{-2em}
}

\maketitle

\begin{abstract}
Integrated terrestrial and non-terrestrial networks (ITNTNs) are regarded as a key architectural paradigm for sixth-generation (6G) wireless systems. This paper investigates a dual-aerial reconfigurable intelligent surface (RIS)-assisted ITNTN, where a terrestrial base station (TBS) and a satellite (SAT) jointly serve terrestrial and satellite users with the aid of an unmanned aerial vehicle (UAV)-mounted RIS and a high-altitude platform (HAP)-mounted RIS. We formulate an average sum-rate maximization problem by jointly optimizing the TBS and SAT precoders, the RIS phase shift matrices, and the three-dimensional trajectories of the UAV and the HAP, subject to transmit power, unit-modulus, and mobility constraints. The resulting optimization problem is highly non-convex due to the strong coupling among the transmit precoders, RIS phase shifts, and aerial platform mobility. To efficiently address this challenge, we propose a block coordinate descent (BCD) framework that integrates weighted minimum mean square error (WMMSE) optimization for precoder design, a manifold-based Riemannian conjugate gradient (RCG) method for RIS phase-shift optimization, and successive convex approximation (SCA) for trajectory optimization. The proposed algorithm is shown to converge to a stationary point. The simulation results show that the proposed joint design achieves an approximately $7.05 \%$ higher average sum-rate compared to the random RIS scheme, highlighting the effectiveness of dual-aerial RIS deployment and joint communication-mobility optimization in ITNTNs.

\end{abstract}
\vspace{-0.5em}
\vspace{-1em}

\section{Introduction}
Sixth-generation (6G) wireless communication systems are envisioned to fundamentally transform network architectures and enabling technologies by supporting global, seamless, and reliable connectivity \cite{11029408}. In this context, integrated terrestrial and non-terrestrial networks (ITNTNs) have emerged as a promising paradigm to complement conventional terrestrial infrastructure by incorporating non-terrestrial platforms such as satellites, high-altitude platforms (HAP), and unmanned aerial vehicles (UAV) \cite{10464935, 10082865,9782683}. By jointly exploiting terrestrial and non-terrestrial resources, ITNTNs can extend coverage to remote and underserved areas in dense urban environments, and improve network resilience in emergency and disaster scenarios \cite{10179219}.
Despite their advantages, ITNTNs face several critical challenges that require careful consideration, including dynamic resource allocation, interference management, coordinated spectrum utilization, and efficient network coordination among platforms operating at different altitudes levels \cite{9861699}. 
\par Reconfigurable intelligent surfaces (RIS), composed of nearly passive and programmable electromagnetic elements, have recently attracted significant attention due to their ability to manipulate wireless propagation environments by dynamically adjusting the phase and amplitude of reflected signals \cite{10988830}. By intelligently configuring RIS phase shifts, wireless channel conditions can be enhanced without relying on additional active radio-frequency chains. Integrating RISs with UAV as relays utilizes the mobility of UAV for flexible RIS deployment and utilizes RIS passive elements for signal reflection and regulation without additional power supply \cite{10938174}. Compared with conventional UAV relays or static RIS-assisted systems, UAV-mounted RIS (UAV-RIS) architectures can achieve improved spectral efficiency with reduced energy consumption \cite{9797690,9741782}.
In parallel, HAP-mounted RIS (HAP-RIS) has recently emerged as a complementary aerial RIS solution. Due to their quasi-stationary operation at high altitudes and wide coverage footprint, HAP-RIS platforms offer enhanced link stability and persistent large-scale coverage, which makes them particularly attractive for satellite-assisted communications \cite{9518388,10186454}. Although UAV-RIS emphasizes mobility and local adaptability, HAP-RIS provides long-term coverage continuity and reduced signaling overhead. The joint deployment of RIS on both UAV and HAP, therefore, enables a dual-aerial RIS architecture that can simultaneously support flexible local coverage enhancement and stable long-range signal reflection within ITNTNs.  

\begin{table*}[t]
    \centering
    \begin{threeparttable} 
    \caption{Comparison of other works with the proposed work}
    \label{tab:ref_comparision}
    \begin{tabular}{|c|c|c|c|c|c|c|}
    \hline
      \diagbox[width=8em,height=3em]{\textbf{References}}{\textbf{Parameter}} & \makecell{\textbf{Beamforming}\\\textbf{Optimization}} & \makecell[c]{\textbf{UAV}\\\textbf{Optimization}} & \makecell[c]{\textbf{HAP}\\\textbf{Optimization}} & \makecell[c]{\textbf{Optimization}\\\textbf{Objective}} & \makecell[c]{\textbf{Optimization}\\\textbf{Method}} & \textbf{Network type}  \\ \hline
     Ref. \cite{10938174} & \makecell[c]{Both active and\\passive beamforming} & \makecell[c]{Fixed line\\trajectory} & - & \makecell[c]{Sum-rate\\maximization} & FP-manifold & \makecell[c]{Terrestrial\\network}  \\ \hline
     Ref. \cite{9797690} & \makecell[c]{Both active and\\passive beamforming} & \makecell[c]{2D UAV\\trajectory} & - & \makecell[c]{Energy efficiency\\maximization} & \makecell[c]{SDR, SROCR \\and SCA} & \makecell[c]{Terrestrial\\network} \\ \hline
     Ref. \cite{10494543} & \makecell[c]{Both active and\\passive beamforming} & \makecell[c]{2D UAV\\trajectory} & - & \makecell[c]{Secrecy-rate\\maximization} & \makecell[c]{S-procedure,\\ SCA and DC} & \makecell[c]{Terrestrial\\network} \\ \hline
     Ref. \cite{10526452} & \makecell[c]{Both active and\\passive beamforming} & Not optimized & - & \makecell[c]{Weighted sum-rate\\maximization} & RoBl & \makecell[c]{Terrestrial\\network} \\ \hline
     Ref. \cite{11008723} & \makecell[c]{Both active and\\passive beamforming} & \makecell[c]{3D UAV\\trajectory} & - & \makecell[c]{Weighted sum-rate\\maximization} & DDQN & \makecell[c]{Terrestrial\\network} \\ \hline
     Ref. \cite{10972091} & \makecell[c]{Both active and\\passive beamforming} & \makecell[c]{3D UAV\\trajectory} & - & \makecell[c]{Sum-rate maximization,\\UAV energy minimization} & SD3 and GNN & \makecell[c]{Terrestrial\\network} \\ \hline
     Ref. \cite{11212820} & \makecell[c]{Both active and\\passive beamforming} & \makecell[c]{Euler angles-based\\UAV control scheme} & - & \makecell[c]{Sum-rate\\maximization} & \makecell[c]{SAC-PER} & \makecell[c]{Terrestrial\\network} \\ \hline
     Ref. \cite{11155890} & \makecell[c]{Both active and\\passive beamforming} & \makecell[c]{3D UAV\\trajectory} & - & \makecell[c]{Maximizing the\\minimum ergodic rate} & \makecell[c]{Alternating\\optimization} & \makecell[c]{Non-Terrestrial\\network} \\ \hline
    Ref. \cite{10184499} & - & - & \makecell[c]{Circular\\trajectory} & \makecell[c]{Channel gain\\maximization} & \makecell[c]{Pareto optimal\\ method} & \makecell[c]{Non-Terrestrial\\network} \\ \hline
    \textbf{Proposed work} & \makecell[c]{Both active and\\passive beamforming} & \makecell[c]{3D UAV\\trajectory} & \makecell[c]{3D HAP\\trajectory} & \makecell[c]{Average sum-rate\\maximization} & \makecell[c]{WMMSE, manifold-\\based RCG and SCA} & ITNTN \\ \hline 
    \end{tabular}
    \end{threeparttable}
    \vspace{-1.25em}
\end{table*}

\vspace{-1em}
\subsection{Related Work}
RIS-assisted communication systems have been extensively investigated in recent years. Early works primarily focused on fixed terrestrial RIS deployments, where transmit beamforming and RIS phase shifts are jointly optimized to enhance coverage and spectral efficiency under static channel conditions. Although these studies demonstrate the potential of RIS-assisted communications, their applicability is limited in highly dynamic non-terrestrial environments.
To improve deployment flexibility, UAV-RIS systems have received increasing attention. Existing UAV-RIS studies mainly consider joint optimization of transmit beamforming, RIS phase shifts, and UAV trajectories for terrestrial communication scenarios. Various performance objectives have been explored, including sum-rate maximization, secure communication, robustness to channel uncertainty, and energy efficiency, using techniques such as block coordinate descent (BCD), successive convex approximation (SCA), and learning-based methods \cite{11124573,10494543,10526452,11008723,10972091,11155890,11212820}. However, these studies typically focus on a single aerial RIS platform and do not consider the coexistence of terrestrial and satellite users within an integrated network framework.
More recently, HAP-RIS systems have been studied as an alternative aerial RIS solution. The existing literature on HAP-RIS mainly investigates coverage enhancement, link budget analysis, or RIS phase shift optimization for terrestrial or satellite communications. These studies indicate that HAP-RIS can achieve better performance compared to UAV-RIS in large-scale coverage scenarios and is particularly suitable for satellite-assisted communications \cite{10184499,10538458,10024826}. Nevertheless, most of the previous works assumes fixed or predictable HAP positions and does not consider joint optimization of beamforming, RIS configuration, and aerial mobility.
\par A few recent studies have explored hierarchical or multi-layer aerial architectures involving both UAV and HAP, mainly from a network architecture or coverage perspective \cite{11274933,10445520}. However, these works focus on user association or coverage partitioning and do not address joint physical-layer optimization. In particular, the strong coupling between terrestrial-satellite beamforming, RIS phase shifts, and the three-dimensional (3D) mobility of both UAV and HAP has not been explicitly investigated in the current literature, as shown in Table~\ref{tab:ref_comparision}.

\vspace{-1em}
\subsection{Motivation and Contributions}
Despite the promising potential of aerial RIS-assisted ITNTN, the joint design of dual-aerial RIS-assisted ITNTNs remains largely unexplored. Most existing studies focus on a single aerial RIS, e.g., either a UAV-RIS or a HAP-RIS, and typically assume fixed aerial platforms, independently optimize active and passive beamforming, or simplified channel models that ignore the complex interconnection between terrestrial and satellite links. As a result, the strong coupling among terrestrial-satellite precoders, dual RIS phase shifts, and aerial platform mobility are not fully exploited, leading to suboptimal system performance. This observation motivates the development of a unified and tractable optimization framework that jointly designs communication and mobility in dual-aerial RIS-assisted ITNTN.
\par In this paper, we investigate a novel dual-aerial RIS-assisted integrated terrestrial and non-terrestrial network architecture, where a terrestrial base station (TBS) and a satellite (SAT) simultaneously serve terrestrial and satellite users with the assistance of a UAV-RIS and a HAP-RIS. We formulate an average sum-rate maximization problem that jointly optimizes the transmit precoders at TBS and SAT, subject to power budgets, the RIS phase shift matrices of the UAV-RIS and the HAP-RIS subject to unit-modulus constraints, and the 3D trajectories of the UAV and the HAP under practical mobility constraints. The resulting problem is highly non-convex due to the coupled variables, non-convex rate expressions, unit-modulus constraints, and trajectory-dependent channel variations.
\par The main contributions of this work are summarized as
\begin{itemize}
    \item We establish a comprehensive system model for a dual-aerial RIS-assisted ITNTN, where a UAV-RIS and a HAP-RIS jointly enhance terrestrial and satellite communications. The proposed model explicitly captures heterogeneous propagation characteristics, 3D mobility of aerial platforms, and cooperative reflection effects from dual RISs, providing a general framework that reduces to single aerial RIS-assisted networks as special cases.
    \item We formulated a average sum-rate maximization problem that jointly optimizes TBS-SAT precoders, UAV-RIS and HAP-RIS phase shifts, and UAV/HAP 3D trajectories, explicitly accounting for the strong coupling among precoders, RIS reflection coefficients, and aerial platform positions.
    \item To tackle the resulting highly non-convex problem, we develop an efficient BCD-based algorithmic framework. In each iteration, the transmit precoding vectors are optimized using the weighted minimum mean square error (WMMSE) algorithm, the unit-modulus constrained RIS phase shift matrices are optimized using the manifold-based Riemannian conjugate gradient (RCG) method, and the non-convex UAV and HAP trajectory subproblems are handled using SCA techniques. The proposed algorithm monotonically converges to a stationary point of the original problem.
    \item Extensive Monte Carlo simulations validate that the proposed algorithm converges in a few iterations. The results demonstrate that the proposed joint design achieves significant sum-rate gains over benchmark schemes employing no RIS, random RIS, or fixed trajectories. These performance improvements are observed across a wide range of system parameters, including convergence behavior, transmit power budget, network density, flight period, noise power, and RIS configurations. The results highlight the effectiveness and robustness of the proposed solution.

\end{itemize}
\vspace{-0.75em}
\subsection{Organization and Notations}
\textit{Organization:} The remainder of this paper is organized as follows. Section \ref{ref:System Model} introduces the system model of dual-aerial RIS-assisted ITNTN. In Section \ref{ref:Problem formulation}, we formulate the joint optimization problem. Then, Section \ref{ref:Proposed solution} presents the proposed BCD-based solution framework. The simulation results are provided in Section \ref{ref:Simulation results} to validate the effectiveness of the proposed design. Finally, Section \ref{ref:Conclusion} concludes the paper.

\textit{Notations:} Scalars, vectors, and matrices are denoted by $x$, $\mathbf{x}$, and $\mathbf{X}$, respectively. $\mathbb{C}^{M\times N}$ denotes the set of complex-valued matrices of size $M \!\times\! N$. $\mathbf{I}_N$ denote the identity matrix of size $N\! \times\! N$, respectively.
The operators $(\cdot)^T$ and $(\cdot)^H$ represent the transpose and the Hermitian transpose, respectively. The symbols $\circ$ and $\otimes$ denote the Hadamard product and the Kronecker product, respectively.
The complex circularly symmetric Gaussian distribution with mean $\mu$ and variance $\sigma^2$ is denoted by $\mathcal{CN}(\mu,\sigma^2)$. The absolute value of a scalar and the Euclidean norm of a vector are denoted by $|\cdot|$ and $\|\cdot\|$, respectively. $\mathrm{diag}(\cdot)$ denotes a diagonal matrix. $\Re\{\cdot\}$ denotes the real part of a complex number, $\iota \triangleq \sqrt{-1}$ denotes the imaginary unit, and $\mathbb{E}[\cdot]$ denotes the expectation operator.

\vspace{-0.75em}
\section{System Model} \label{ref:System Model}
\begin{figure}[t]
    \centering
    \includegraphics[scale=0.13]{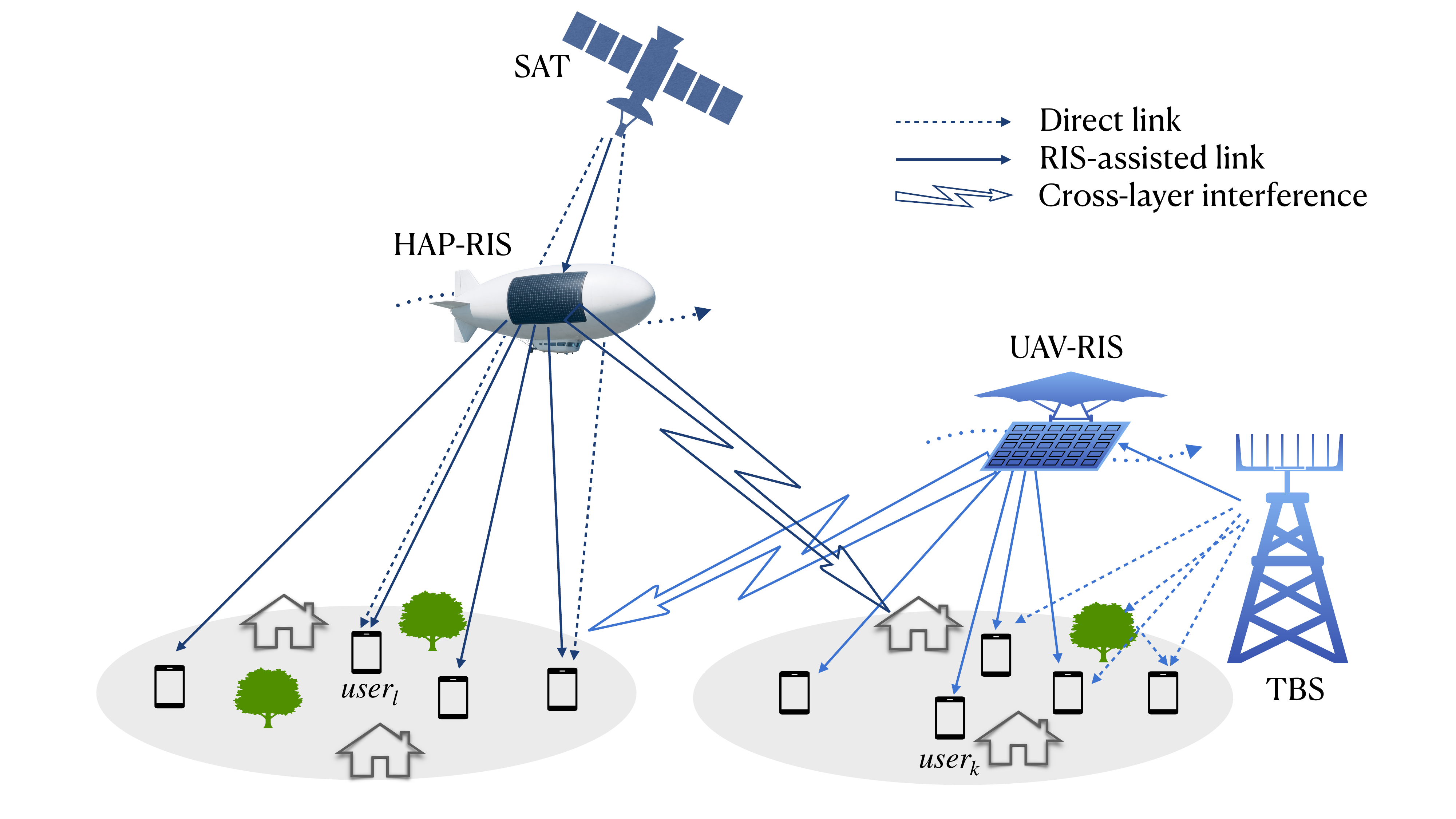}
    \caption{System Model of Aerial RIS-assisted Integrated Terrestrial and Non-terrestrial Network.}
    \label{fig:1}
\end{figure}

We consider an ITNTN in which the TBS is equipped with a uniform linear array (ULA) comprising $M_b$ antennas with inter-element spacing $d \!=\! \lambda/2$, where $\lambda\!=\!c/f_c$ denotes the carrier wavelength of the terrestrial link, $c$ denote the speed of light, and $f_c$ denotes the carrier frequency of the terrestrial link. The TBS servers a region containing $K$ users indexed by $k \in  \{1,\dots,K\}$. In parallel, the SAT is equipped with a uniform planar array (UPA) comprising $M_s \!=\! N_{s,x}\!\times \!N_{s,y}$ antennas with inter-element spacing $\check{d} \!=\! \check{\lambda}/2$, where $\check{\lambda}\!=\!c/\check{f}_c$ denotes the carrier wavelength, and $\check{f}_c$ denotes the carrier frequency of the satellite link. The SAT serves a separate region containing $L$ users indexed by $l \in \{1,\dots,L\}$. 
An UAV-mounted RIS (UAV-RIS) is equipped with a UPA of $N_U = N_x \times N_y$ passive reflecting elements assists the TBS transmissions, while an HAP-mounted RIS (HAP-RIS) with a UPA of $N_H = N_{h,x}\times N_{h,y}$ passive reflecting elements assists the SAT transmissions. Both the UAV and the HAP are mobile, and their trajectories are jointly optimized to improve the network performance.
A data frame of duration $T$ is divided into $N$ time slots of equal length, each of duration $\delta = \frac{T}{N}$. The slot duration $\delta$ is chosen such that UAV-RIS, HAP-RIS, and user locations remain quasi-static within each time slot, while avoiding excessive control and computational overhead due to frequent updates \cite{11155890}. As illustrated in Fig.~\ref{fig:1}, all communication nodes are located in a 3D Cartesian coordinate system. 
\vspace{-0.75em}
\subsection{Channel Model}
Channel modeling plays a key role in determining the performance and reliability of wireless communication systems. In this section, we present the channel models associated with terrestrial, aerial, and satellite links of the integrated network. 
 
\subsubsection{Terrestrial-to-user Link}
Consider a TBS located at $\mathbf{q}_{_{TBS}}= (x_{_{TBS}}, y_{_{TBS}}, 0)$, which provides communication services to $K$ TBS users located at $\mathbf{q}_{_k} = (x_{_k}, y_{_k}, 0), \forall k$. Due to potential blockages caused by surrounding obstacles, the quality of the direct TBS-to-user link may be severely degraded. To improve communication reliability and coverage, a UAV-RIS is deployed to establish an additional reflective link between the TBS and the users. The 3D position of the UAV-RIS during the $n$-th time slot is denoted by $\textbf{q}_{_U}[n] = \big(x_{_U}[n],y_{_U}[n],z_{_U}[n]\big)$.

By using the Rician fading model, the direct TBS-to-user $k$ channel $\mathbf{h}_{TBS,k} \in \mathbb{C}^{1\times M_b}$ is modeled as \cite{11212820}
\begin{align}
\mathbf{h}_{TBS,k}
= \sqrt{\beta_{TBS,k}} \Bigg(&
\sqrt{\frac{\kappa_{_{TBS,k}}}{\kappa_{_{TBS,k}}+1}}\,
\mathbf{h}_{TBS,k}^{\mathrm{LoS}}\nonumber \\
&\quad +
\sqrt{\frac{1}{\kappa_{_{TBS,k}}+1}}\,
\mathbf{h}_{TBS,k}^{\mathrm{NLoS}}
\Bigg),
\end{align}
where $\kappa_{_{TBS,k}}$ is the Rician factor, and $\beta_{TBS,k}$ denotes the large-scale fading component of the direct link, given by 
\begin{equation}
    \beta_{TBS,k} = \beta_o d_{TBS,k}^{-\alpha_{_{TBS,k}}},
\end{equation}
where $\beta_o$ denotes the reference channel gain at a distance of $d = 1$ m, $d_{TBS,k}=\|\mathbf{q}_{_{TBS}}-\mathbf{q}_{_k}\|$ is the TBS-to-user $k$ distance, and $\alpha_{_{TBS,k}}$ denotes the path loss exponent. The NLoS component $\mathbf{h}_{TBS,k}^{\mathrm{NLoS}}$ follows a circularly symmetric complex Gaussian distribution $\mathcal{CN}(0,\mathbf{I}_{ M_b})$. Let $\psi_{_{TBS}}^k$ and $\phi_{_{TBS}}^k$ denote the azimuth and elevation angle of arrival (AoA) at user $k$. The LoS component of the direct link is given by \cite{11008723}
\begin{equation}
    \mathbf{h}_{TBS,k}^{\mathrm{LoS}} = \mathbf{a}^H_{_{TBS}}(\psi_{_{TBS}}^k,\phi_{_{TBS}}^k),
\end{equation}
The TBS ULA steering vector is given by
\begin{align}
    \mathbf{a}_{_{TBS}}(\psi,\phi) = \frac{1}{\sqrt{M_b}}\Big[ 1, &e^{-\iota2\pi\frac{d}{\lambda}\cos\phi\cos\psi},
    \dots,\nonumber\\&\hspace{2em} e^{-\iota2\pi\frac{(M_b-1)d}{\lambda}\cos\phi\cos\psi}
    \Big]^T,
\end{align}
where $\psi$ and $\phi$ denote the azimuth and elevation angles, respectively. For the RIS-assisted link, the communication path from the TBS to user $k$ consists of two cascaded sub-links: (i) the TBS-to-UAV-RIS link and (ii) the UAV-RIS-to-user link. 
\par The channel from TBS to UAV-RIS $\mathbf{H}_{TBS,U} \in \mathbb{C}^{N_U\times M_b}$ is modeled as \cite{11212820}
\begin{align}
\mathbf{H}_{TBS,U} 
= \sqrt{\beta_{TBS,U}} \Bigg(&
\sqrt{\frac{\kappa_{_{TBS,U}}}{\kappa_{_{TBS,U}}+1}}\,
\mathbf{H}_{TBS,U}^{\mathrm{LoS}} \nonumber\\
&\hspace{2em} +
\sqrt{\frac{1}{\kappa_{_{TBS,U}}+1}}\,
\mathbf{H}_{TBS,U}^{\mathrm{NLoS}}
\Bigg),
\end{align}
where $\kappa_{_{TBS,U}}$ is the Rician factor, and $\beta_{TBS,U}$ denotes the large-scale fading component, given by 
\begin{equation}
    \beta_{TBS,U} = \beta_o d_{TBS,U}^{-\alpha_{_{TBS,U}}},
\end{equation}
where $d_{TBS,U}=\|\mathbf{q}_{_U}[n]-\mathbf{q}_{_{TBS}}\|$ denotes the distance between TBS and UAV-RIS, and $\alpha_{_{TBS,U}}$ denotes the path loss exponent. The NLoS component $\mathbf{H}_{TBS,U}^{\mathrm{NLoS}}$ entries are i.i.d. circularly symmetric complex Gaussian distribution with zero mean and unit variance.
The UAV-RIS UPA steering vector is \cite{11008723}
\begin{equation}
    \mathbf{a}_{_U}(\psi,\phi) = \frac{1}{\sqrt{N_U}}\big(\mathbf{a}_x(\psi,\phi)\otimes\mathbf{a}_y(\psi,\phi)\big),
\end{equation}
where
\begin{align}
\mathbf{a}_{x}(\psi,\phi)\! &=\! \frac{1}{\sqrt{N_x}}\Big[\! 1, e^{-\iota2\pi\frac{d_x}{\lambda}\cos\phi\cos \psi},\nonumber\\& \hspace{5em}
\dots, e^{-\iota2\pi\frac{(N_x-1)d_x}{\lambda}\cos\phi\cos \psi}
\!\Big]^T, \\
\mathbf{a}_{y}(\psi,\phi)\! &=\! \frac{1}{\sqrt{N_y}}\Big[\! 1, e^{-\iota2\pi\frac{d_y}{\lambda}\cos\phi\sin \psi},\nonumber \\& \hspace{5em}
\dots, e^{-\iota2\pi\frac{(N_y-1)d_y}{\lambda}\cos\phi\sin \psi}
\!\Big]^T,
\end{align}
Let $(\psi_{_U},\phi_{_U})$ denote the azimuth and elevation AoA at the UAV-RIS from the TBS, and $(\psi_{_{TBS}}^U,\phi_{_{TBS}}^U)$ denote the angle of departure (AoD) at the TBS. The LoS component of TBS-to-UAV-RIS link is given by\cite{11008723}
\begin{equation}
    \mathbf{H}_{TBS,U}^{\mathrm{LoS}} = \mathbf{a}_{_U}(\psi_{_U},\phi_{_U})\mathbf{a}_{_{TBS}}^H(\psi_{_{TBS}}^U,\phi_{_{TBS}}^U).
\end{equation}
Similarly, the channel between the UAV-RIS and the user $k$ $\mathbf{h}_{U,k} \in \mathbb{C}^{N_U\times 1}$ is modeled as \cite{11212820}
\begin{align}
\mathbf{h}_{U,k}
= \sqrt{\beta_{U,k}} \Bigg(
\sqrt{\frac{\kappa_{_{U,k}}}{\kappa_{_{U,k}}\!+\!1}}
\mathbf{h}_{U,k}^{\mathrm{LoS}} 
 +
\sqrt{\frac{1}{\kappa_{_{U,k}}\!+\!1}}
\mathbf{h}_{U,k}^{\mathrm{NLoS}}
\Bigg),
\end{align}
where $\kappa_{_{U,k}}$ is the Rician factor, and $\beta_{U,k}$ denotes the large-scale fading component, given by 
\begin{equation}
    \beta_{U,k} = \beta_o d_{U,k}^{-\alpha_{_{U,k}}},
\end{equation}
where $d_{U,k}\!=\!\|\mathbf{q}_{_U}[n]\!-\!\mathbf{q}_{_k}\|$ is the distance between UAV-RIS and the user $k$, and $\alpha_{_{U,k}}$ denotes the path loss exponent. The NLoS component $\mathbf{h}_{U,k}^{\mathrm{NLoS}}$ follows a circularly symmetric complex Gaussian distribution $\mathcal{CN}(0,\mathbf{I}_{N_U \times 1})$.
Let $(\psi_{_U}^k,\phi_{_U}^k)$ be the azimuth and elevation AoD from the UAV-RIS to user $k$. The LoS component of UAV-RIS-to-user $k$ is given by \cite{11008723}
\begin{equation}
    \mathbf{h}_{U,k}^{\mathrm{LoS}} = \mathbf{a}_U(\psi_{_U}^k,\phi_{_U}^k).
\end{equation}
Finally, at the $n$-th time slot, the effective channel between TBS and the user $k$ can be expressed as
\begin{equation}
    \mathbf{h}_{tr,k}[n] = \mathbf{h}_{TBS,k}+\mathbf{h}_{U,k}^H[n] \boldsymbol{\Theta}_U[n] \mathbf{H}_{TBS,U}[n],
\end{equation}
where UAV-RIS phase shift matrix is $\boldsymbol{\Theta}_U \!=\! \text{diag}(e^{\iota\theta_1},\dots,e^{\iota\theta_{N_U}}) \\\in \mathbb{C}^{N_U \times N_U}$, with $\theta_i \in [0,2\pi)$ denotes the adjustable phase shift of the $i$-th reflecting element of the UAV-RIS.\vspace{-0.5em}
\subsubsection{Satellite-to-user Link}
Consider a SAT located at $ \mathbf{q}_{_{SAT}} = (x_{_{SAT}}, y_{_{SAT}}, H_{{SAT}})$, where $H_{SAT}$ is the altitude of the satellite. The satellite servers $L$ SAT users located at $\mathbf{q}_{_l} = (x_{_l}, y_{_l}, 0), \forall \; l$.  Due to severe free-space path loss and weather-induced impairments, the direct SAT-to-user links are generally weak. To improve link reliability and coverage, a HAP-RIS is deployed to establish an additional reflective link between the SAT and users. The 3D position of the HAP-RIS during the $n$-th time slot is denoted by $\textbf{q}_{_H}[n] = \big(x_{_H}[n],y_{_H}[n],z_{_H}[n]\big)$. The direct SAT-to-user $l$ channel is modeled as a Rician fading and is given by \cite{11155890,9628071}
\begin{align}
\mathbf{h}_{SAT,l}
\!=\! \mathbb{L}_{SAT,l} \!\Bigg(\!\!
\sqrt{\!\frac{\kappa_{_{SAT,l}}}
              {\kappa_{_{SAT,l}}\!\!+\!1}}
   \mathbf{h}_{SAT,l}^{\mathrm{LoS}}  \!\! + \!\! 
  \sqrt{\!\frac{1}{\kappa_{_{SAT,l}}\!\!+\!1}\!}
   \mathbf{h}_{SAT,l}^{\mathrm{NLoS}}\!\!
\Bigg)\!,
\end{align}
where $\kappa_{_{SAT,l}}$ denotes the Rician factor and $\mathbb{L}_{SAT,l}$ represents the large-scale fading gain, given by
\begin{equation}
\mathbb{L}_{SAT,l} =\sqrt{\left(\frac{\check{\lambda}}{4\pi d_{SAT,l}}
\right)^{2}G_{SAT}G_lr_s
},
\end{equation}
where $d_{SAT,l} = \|\mathbf{q}_{_{SAT}}-\mathbf{q}_{_l}\|$ denotes the SAT-to-user $l$ distance, while $G_{SAT}$ and $G_l$ denote the antenna gain of the SAT and the user, respectively.  $r_s$ represents the rain attenuation factor, modeled as $\ln(r_s^{\mathrm{dB}})\sim\mathcal{N}(\mu_r,\sigma_r^2)$ \cite{3GPP_TR_38_901}. The NLoS component follows $\mathbf{h}_{SAT,l}^{\mathrm{NLoS}} \sim \mathcal{CN}(0,\mathbf{I}_{ M_s})$. 
The SAT UPA responses are given by \cite{11155890,9628071}
\begin{equation}
    \mathbf{a}_{_{SAT}}(\psi,\phi) = \frac{1}{\sqrt{M_s}}\big(\mathbf{a}_{_{SAT}}^x(\psi,\phi)\otimes\mathbf{a}_{_{SAT}}^y(\psi,\phi)\big),
\end{equation}
where
\begin{align}
&\mathbf{a}_{_{SAT}}^{x}(\psi,\phi)
= \frac{1}{\sqrt{M_{s,x}}}
\Big[
 1,
e^{-\iota2\pi\frac{\check{d}_x}{\check{\lambda}}\cos\phi\cos\psi},  \nonumber\\
& \hspace{9em}\dots,
e^{-\iota2\pi\frac{(M_{s,x}-1)\check{d}_x}{\check{\lambda}}
     \cos\phi\cos\psi}
\Big]^{T},\\
&\mathbf{a}_{_{SAT}}^{y}(\psi,\phi)
= \frac{1}{\sqrt{M_{s,y}}}
\Big[
 1,
e^{-\iota2\pi\frac{\check{d}_y}{\check{\lambda}}\cos\phi\sin\psi},\nonumber\\ &\hspace{9em} \dots,e^{-\iota2\pi\frac{(M_{s,y}-1)\check{d}_y}{\check{\lambda}}\cos\phi\sin\psi}
\Big]^{T},
\end{align}
Here, $M_{s,x}$ and $M_{s,y}$ are the numbers of SAT UPA elements along the $x$ and $y$-axes, respectively, and $\check{d}_x$ and $\check{d}_y$ are the corresponding antenna spacings. Let $\psi_{sat}^l$ and $\phi_{sat}^l$ denote the azimuth and elevation angles of user $l$, respectively. The LoS component corresponds to the SAT UPA steering vector, given by \cite{11155890,9628071}
\begin{equation}
\mathbf{h}_{SAT,l}^{\mathrm{LoS}} = \mathbf{a}_{_{SAT}}^H(\psi_{sat}^l,\phi_{sat}^l).
\end{equation}
The SAT-to-HAP-RIS channel can be modeled as \cite{11155890,9628071}
\begin{align}
\mathbf{H}_{SAT,H}
=
\mathbb{L}_{SAT,H}
\Bigg(
& \sqrt{\frac{\kappa_{_{SAT,H}}}
              {\kappa_{_{SAT,H}}+1}}
\, \mathbf{H}_{SAT,H}^{\mathrm{LoS}}\nonumber \\
& +
\sqrt{\frac{1}{\kappa_{_{SAT,H}}+1}}
\, \mathbf{H}_{SAT,H}^{\mathrm{NLoS}}
\Bigg),
\end{align}
where
\begin{equation}
\mathbb{L}_{SAT,H} = \sqrt{\left(\frac{\check{\lambda}}{4\pi d_{SAT,H}}\right)^2G_{SAT}G_{HAP}},
\end{equation}
with $d_{SAT,H}=\|\mathbf{q}_{_{SAT}}-\mathbf{q}_{_{H}}[n]\|$. The NLoS term $\mathbf{H}_{SAT,H}^{\mathrm{NLoS}}$ entries are i.i.d. circularly symmetric complex Gaussian distribution with zero mean and unit variance. The HAP-RIS UPA  response is given by \cite{11155890,9628071}
\begin{align}
    &\mathbf{a}_{_H}(\psi,\phi) = \frac{1}{\sqrt{N_H}}\big(\mathbf{a}_{_H}^x(\psi,\phi)\otimes\mathbf{a}_{_H}^y(\psi,\phi)\big),\\
&\mathbf{a}_{_H}^{x}(\psi,\phi)
= \frac{1}{\sqrt{N_{h,x}}}
\Big[
 1,
e^{-\iota2\pi\frac{\check{d}_{h,x}}{\check{\lambda}}\cos\phi\cos\psi},  \nonumber\\
&\hspace{5em} \dots,\;
e^{-\iota2\pi\frac{(N_{h,x}-1)\check{d}_{h,x}}{\check{\lambda}}
     \cos\phi\cos\psi}
\Big]^{T},
\end{align}
and
\begin{align}
\mathbf{a}_{_H}^{y}(\psi,\phi)
&= \frac{1}{\sqrt{N_{h,y}}}
\Big[
 1,
e^{-\iota2\pi\frac{\check{d}_{h,y}}{\check{\lambda}}\cos\phi\sin\psi},\nonumber\\ &\hspace{5em} \dots,e^{-\iota2\pi\frac{(N_{h,y}-1)\check{d}_{h,y}}{\check{\lambda}}\cos\phi\sin\psi}
\Big]^{T},
\end{align}
Here, $N_{h,x}$ and $N_{h,y}$ are the numbers of HAP-RIS UPA elements along the $x$ and $y$-axes, respectively, and $\check{d}_{h,x}$ and $\check{d}_{h,y}$ are the corresponding antenna spacings. Let $\psi_{_H}$ and $\phi_{_H}$ denote the azimuth and elevation AoA at the HAP-RIS from the SAT, and $\psi_{sat}^h$ and $\phi_{sat}^h$ denote the AoD at the SAT.
The LoS component of SAT-to-HAP-RIS link is given by \cite{11155890,9628071}
\begin{equation}
\mathbf{H}_{SAT,H}^{\mathrm{LoS}} = \mathbf{a}_{_H}(\psi_{_H},\phi_{_H})\mathbf{a}_{_{SAT}}^H(\psi_{sat}^h,\phi_{sat}^h).
\end{equation}
The HAP-RIS-to-user $l$ channel is given by \cite{11155890,9628071}
\begin{align}
\mathbf{h}_{H,l}
=\mathbb{L}_{H,l}
\Bigg(\sqrt{\frac{\kappa_{_{H,l}}}
    {\kappa_{_{H,l}}+1}} \mathbf{h}_{H,l}^{\mathrm{LoS}} +
\sqrt{\frac{1}{\kappa_{_{H,l}}+1}} \mathbf{h}_{H,l}^{\mathrm{NLoS}}
\Bigg),
\end{align}
where \vspace{-1.25em}
\begin{equation}
\mathbb{L}_{H,l}
=
\sqrt{
\left(
\frac{\check{\lambda}}{4\pi d_{H,l}}
\right)^{2}
G_{HAP}G_lr_s
},
\end{equation}
with $d_{H,l} = \|\mathbf{q}_{_H}[n]-\mathbf{q}_{_l}\|$ denotes the HAP-RIS-to-user $l$ distance. The NLoS component satisfies $\mathbf{h}_{H,l}^{\mathrm{NLoS}} \sim \mathcal{CN}(0,\mathbf{I}_{N_{H}})$, while the LoS component is \cite{11155890,9628071}
\begin{align}
\mathbf{h}_{H,l}^{\mathrm{LoS}}
= \mathbf{a}_{_H}(\psi_{_H}^l,\phi_{_H}^l),
\end{align}
where $(\psi_{_H}^l,\phi_{_H}^l)$ represents the azimuth and elevation AoD from HAP-RIS to user $l$.  

Consequently, the effective channel between the SAT and the user $l$ during the time slot $n$ can be expressed as
\begin{equation}
\mathbf{h}_{s,l}[n]
=
\mathbf{h}_{SAT,l}
+
\mathbf{h}_{H,l}^{H}[n]
\boldsymbol{\Theta}_H[n]
\mathbf{H}_{SAT,H}[n],
\end{equation}
where the HAP-RIS phase shift matrix is given by \(\boldsymbol{\Theta}_{H} = \mathrm{diag}\!\left(e^{\iota\theta_{1}}, e^{\iota\theta_{2}}, \dots, e^{\iota\theta_{N_\mathbb{H}}}\right)\in\mathbb{C}^{N_H \times N_H}\), with \(\theta_{j} \in [0,2\pi)\) denotes the adjustable phase shift of the \(j\)-th reflecting element of the HAP-RIS. \vspace{-2em}
\subsection{Input-output relations}
We assume linear precoding at the TBS and the SAT and single data stream per user.
At time slot $n$, the TBS transmits
\begin{equation}
\mathbf{x}_b[n] = \sum\nolimits_{k=1}^{K} \mathbf{w}_{b,k}[n]\, s_{b,k}[n],
\end{equation}
where $\mathbf{w}_{b,k}[n]\in\mathbb{C}^{M_b\times 1}$ is the precoding vector for TBS user $k$
and $s_{b,k}[n]$ is the information symbol with $\mathbb{E}\big[|s_{b,k}[n]|^2\big]=1$.
Similarly, the SAT transmits
\begin{equation}
\mathbf{x}_s[n] = \sum\nolimits_{l=1}^{L} \mathbf{w}_{s,l}[n]\, s_{s,l}[n],
\end{equation}
where $\mathbf{w}_{s,l}[n]\in\mathbb{C}^{M_s\times 1}$ is the precoding vector for SAT user $l$ and $s_{s,l}[n]$ is the information symbol with $\mathbb{E}\big[|s_{s,l}[n]|^2\big]=1$.
\subsubsection{Received signal at TBS user $k$} 
The baseband received signal at TBS user $k$ during the slot $n$ is
\begin{align}
y_{_k}[n] &\!=\! \underbrace{\mathbf{h}_{tr,k}[n]\,\mathbf{w}_{b,k}[n]\,s_{b,k}[n]}_{\text{desired}} +\!\underbrace{\sum\nolimits_{i\neq k}\mathbf{h}_{tr,k}[n]\,\mathbf{w}_{b,i}[n]\,s_{b,i}[n]}_{\text{intra-TBS interference }} \nonumber\\
&\hspace{3em}+\underbrace{\sum\nolimits_{l=1}^{L}\mathbf{g}_{_{{SAT}\to k}}[n]\,\mathbf{w}_{s,l}[n]\,s_{s,l}[n]}_{\text{SAT$ \to$ user $k$ interference}\;}+ n_k[n],
\end{align}
where $n_k[n]\sim\mathcal{CN}(0,\sigma_k^2)$ denotes additive white Gaussian noise (AWGN) at the user $k$ with zero mean and variance $\sigma_k^2$.
For any TBS user $k$, the interference signal originating from the satellite comprises a direct SAT-to-user link and an RIS-reflected SAT-to-HAP-RIS-to-user link. The corresponding cross-tier channel is defined as
\begin{equation}
\mathbf{g}_{_{SAT\to k}}
=
\mathbf{h}_{\mathrm{SAT},k}
+
\mathbf{h}_{H,k}^{H}
\boldsymbol{\Theta}_{H}
\mathbf{H}_{SAT,H},
\label{eq:g_SAT_k}
\end{equation}
where \(\mathbf{h}_{SAT,k}\) denotes the direct SAT-to-user
\(k\) channel, and \(\mathbf{h}_{H,k}\) is the HAP-RIS-to-user $k$ channel. 

The instantaneous signal to interference plus noise ratio (SINR) of the user $k$ is
\begin{equation}
\mathrm{SINR}_{k}[n]
=
\frac{\big|\mathbf{h}_{tr,k}[n]\mathbf{w}_{b,k}[n]\big|^2}
{I_{k}^{\mathrm{intra}}[n] + I_{k}^{\mathrm{cross}}[n] + \sigma_k^2 },
\end{equation}
where 
\begin{align}
    I_{k}^{\mathrm{intra}}[n] &= \sum\nolimits_{i\neq k}\big|\mathbf{h}_{tr,k}[n]\,\mathbf{w}_{b,i}[n]\big|^2, \nonumber \\
    I_{k}^{\mathrm{cross}}[n] &= \sum\nolimits_{l=1}^{L}\big|\mathbf{g}_{_{{SAT}\to k}}[n]\,\mathbf{w}_{s,l}[n]\big|^2.
\end{align}

\subsubsection{Received signal at SAT user $l$}
The received signal at SAT user $l$ is
\begin{align}
y_{_l}[n] &= \underbrace{\mathbf{h}_{s,l}[n]\,\mathbf{w}_{s,l}[n]\,s_{s,l}[n]}_{\text{desired}}\nonumber 
+\underbrace{\sum\nolimits_{j\neq l}\mathbf{h}_{s,l}[n]\,\mathbf{w}_{s,j}[n]\,s_{s,j}[n]}_{\text{intra-SAT interference}}\nonumber \\
&\hspace{3em}+\underbrace{\sum\nolimits_{k=1}^{K}\mathbf{g}_{_{TBS\to l}}[n]\,\mathbf{w}_{b,k}[n]\,s_{b,k}[n]}_{\text{TBS $\to$ user $l$ interference}} + n_{_l}[n],
\end{align}
where $n_{_l}[n]\sim\mathcal{CN}(0,\sigma_l^2)$ denotes AWGN at user $l$ with zero mean and variance $\sigma_l^2$.
The interference caused by TBS to a SAT user \(l\) consists of a direct TBS-to-user link and a reflected TBS-to-UAV-RIS-to-user link. The cross-tier channel is given by
\begin{equation}
\mathbf{g}_{_{TBS\to l}}
=
\mathbf{h}_{TBS,l}
+
\mathbf{h}_{U,l}^{H}
\boldsymbol{\Theta}_{U}
\mathbf{H}_{TBS,U},
\label{eq:g_TBS_l}
\end{equation}
where \(\mathbf{h}_{TBS,l}\) denotes the direct TBS-to-user \(l\)
channel, and \(\mathbf{h}_{U,l}\) is the UAV-RIS-to-user channel.

The SINR at SAT user $l$ is
\begin{equation}
\mathrm{SINR}_{l}[n]
=
\frac{\big|\mathbf{h}_{s,l}[n]\mathbf{w}_{s,l}[n]\big|^2}
{I_{s,l}^{\mathrm{intra}}[n] + I_{s,l}^{\mathrm{cross}}[n] + \sigma_l^2 },
\end{equation}
where 
\begin{align}
    I_{l}^{\mathrm{intra}}[n] & = \sum\nolimits_{j\neq l}\big|\mathbf{h}_{s,l}[n]\mathbf{w}_{s,j}[n]\big|^2, \nonumber\\
    I_{l}^{\mathrm{cross}}[n] & = \sum\nolimits_{k=1}^{K}\big|\mathbf{g}_{_{TBS\to l}}[n]\mathbf{w}_{b,k}[n]\big|^2.
\end{align}
Let $R_k[n]$ and $R_l[n]$ denote the achievable rates of TBS user $k$ and SAT user $l$ at time slot $n$, which are given by
\begin{equation}
    R_k[n]\!=\! \log_2\big(1 \!+\! \mathrm{SINR}_k[n]\big), \;\;
R_l[n] \!=\! \log_2\big(1 \!+\! \mathrm{SINR}_l[n]\big).
\end{equation}
Accordingly, the average sum-rate over a transmission frame consisting of 
$N$ time slots is expressed as
\begin{equation}\label{eq:sum_rate}
    \mathcal{R} = \frac{1}{N}\sum\nolimits_{n=1}^{N} \bigg( \sum\nolimits_{k=1}^{K}  R_k[n] + \sum\nolimits_{l=1}^{L}  R_l[n] \bigg),
\end{equation}
\section{Problem Formulation}\label{ref:Problem formulation}
The primary objective of this work is to maximize the average sum-rate by jointly optimizing the system design variables over the entire frame of $N$ time slots. Consequently, the sum-rate maximization problem can be formulated as
\begin{subequations} \label{OPT}
\begin{align}
\textbf{(P1)}\quad&\max_{\substack{\{\mathbf{w}_{b,k}[n],\mathbf{w}_{s,l}[n]\},\\ \{\boldsymbol{\Theta}_{U}[n],\boldsymbol{\Theta}_{H}[n]\}, \{\mathbf{q}_U[n],\mathbf{q}_{H}[n]\}}}\qquad 
 \mathcal{R} \label{opt:obj} \\
\text{s.t.} \quad 
& \sum\nolimits_{k=1}^{K} \|\mathbf{w}_{b,k}[n]\|^2 \le P_b, \quad \forall n, \label{opt:TBS_power} \\
& \sum\nolimits_{l=1}^{L} \|\mathbf{w}_{s,l}[n]\|^2 \le P_s, \quad \forall n, \label{opt:SAT_power} \\
& |\theta_{U,i}[n]| = 1, \quad i=1,\dots,N_U, \quad \forall n, \label{opt:UAV_RIS_unit} \\
& |\theta_{H,j}[n]| = 1, \quad j=1,\dots,N_H, \quad \forall n, \label{opt:HAP_RIS_unit} \\
& \|\mathbf{q}_{_U}[n+1]-\mathbf{q}_{_U}[n]\| \leq V_{U,max}\cdot \delta \quad \forall n, \label{opt:UAV trajectories}\\
& \|\mathbf{q}_{_H}[n+1]-\mathbf{q}_{_H}[n]\| \leq V_{H,max}\cdot \delta \quad \forall n, \label{opt:HAP trajectories} \\
& z_{_U}^{min} \le z_{_U}[n] \le z_{_U}^{max} \quad \forall n, \label{opt:UAV altitude} \\
& z_{_H}^{min} \le z_{_H}[n] \le z_{_H}^{max} \quad \forall n, \label{opt:HAP altitude} \\
& \mathbf{q}_{_U}[1] = \mathbf{q}_{_U}^{\text{init}}, \quad \mathbf{q}_{_U}[N] = \mathbf{q}_{_U}^{\text{final}}, \label{opt:UAV init-final} \\
& \mathbf{q}_{_H}[1] = \mathbf{q}_{_H}^{\text{init}}, \quad \mathbf{q}_{_H}[N] = \mathbf{q}_{_H}^{\text{final}}, \label{opt:HAP init-final}
\end{align}
\end{subequations}
where $P_b$ and $P_s$ denote the maximum transmit powers of the TBS and SAT,  $\theta_{U,i}[n]$ and $\theta_{H,j}[n]$ represent the UAV-RIS and HAP-RIS phase shifts, $V_{U,max}$ and $V_{H,max}$ are the maximum speeds of UAV and HAP, respectively. 
Constraints (\ref{opt:TBS_power}) and (\ref{opt:SAT_power}) limit the transmit power at the TBS and SAT, respectively. Constraints (\ref{opt:UAV_RIS_unit}) and (\ref{opt:HAP_RIS_unit}) enforce the unit-modulus property required for RIS phase-shift design, while (\ref{opt:UAV trajectories}) and (\ref{opt:HAP trajectories}) specify the mobility limits of the UAV and HAP, respectively. 
Finally, constraint \eqref{opt:UAV altitude} and \eqref{opt:HAP altitude} ensure that the UAV and HAP altitudes remain within a feasible operational range, whereas \eqref{opt:UAV init-final} and \eqref{opt:HAP init-final} specify the initial and final positions of the UAV and HAP, respectively.
Solving the average sum-rate maximization problem \textbf{(P1)} is highly challenging due to the non-convex nature of the objective function, the unit-modulus constraints imposed on the RIS phase shifts, and the intricate coupling among the optimization variables $\{\mathbf{w}_{b,k},\mathbf{w}_{s,l}\},\{  \boldsymbol{\Theta}_{\mathbb{U}},\boldsymbol{\Theta}_{{H}}\}, \{\mathbf{q}_{_U},\mathbf{q}_{_{H}}\}$. Consequently, the resulting optimization problem is intractable, posing significant challenges for the joint design of precoding vectors, RIS phase shifts, and trajectories.
\section{Proposed Solution}\label{ref:Proposed solution}
We employ an iterative BCD framework, where the problem is divided into three subproblems: 1) Optimize TBS and SAT precoders using WMMSE by fixing RIS phases and trajectories, 2) Optimize UAV-RIS and HAP-RIS phase shifts using manifold-based RCG method by fixing trajectories and precoders, and 3) Optimize UAV and HAP trajectories using SCA by fixing precoders and RIS phases. The following are the optimization steps for each of these three subproblems.
\vspace{-1.25em}
\subsection{TBS and SAT Precoder Optimization}
For fixed UAV-RIS and HAP-RIS phase shifts, and trajectories, the corresponding BCD subproblem reduces to the optimization of the transmit precoding vectors at the TBS and the SAT. The precoders $\mathbf{w}_{b,k}$ and $ \mathbf{w}_{s,l}$ are therefore optimized by solving the following sum-rate maximization problem
\begin{subequations}
\begin{align}
\textbf{(P2)}\quad &\max_{\{\mathbf{w}_{b,k},\mathbf{w}_{s,l}\}}\quad
  \sum\nolimits_{k=1}^{K}  R_k + \sum\nolimits_{l=1}^{L}  R_l   \\
&\text{s.t.} \quad \eqref{opt:TBS_power} \;\&\; \eqref{opt:SAT_power}. 
\label{P2}
\end{align}
\end{subequations}
The subproblem \textbf{(P2)} is non-convex due to the coupled SINR expressions in the achievable rates. To efficiently address this, we employ the WMMSE algorithm, which transforms the original sum-rate maximization problem into an equivalent and tractable WMMSE minimization problem by introducing auxiliary equalizer and weight variables \cite{10054084,5756489}.
\subsubsection{WMMSE Reformulation}
By introducing the auxiliary receive equalizers $\{u_k,u_l\}$ and the positive weight variables $\{\omega_k,\omega_l\}$, subproblem \textbf{(P2)} can be equivalently reformulated as the following WMMSE minimization problem,
\begin{subequations}
\begin{align}
 \textbf{(P3)} \quad  &\min_{\substack{\{\mathbf{w}_{b,k},\mathbf{w}_{s,l}\},\\ \{u_k,u_l\},\{\omega_k,\omega_l\}}} 
\quad \sum\nolimits_{k=1}^K\!\check{R}_k +\sum\nolimits_{l=1}^L\check{R}_l \\
&\text{s.t.} \quad 
\eqref{opt:TBS_power} \;\&\; \eqref{opt:SAT_power}, 
\end{align}
\end{subequations}
where 
$\check{R}_k = \omega_ke_k\!-\!\log\omega_k$, $\check{R}_l = \omega_le_l\!-\!\log\omega_l$,
and the per-user mean-square errors (MSEs) are given by
\begin{subequations}
\begin{align}
    e_k & = |u_k|^2\mathrm{T}_k- 2\Re\{u_k\mathbf{h}_{tr,k}\mathbf{w}_{b,k}\} +1, \label{eq:TBS MSE}\\
    e_l &= |u_l|^2\mathrm{T}_l- 2\Re\{u_l\mathbf{h}_{s,l}\mathbf{w}_{s,l}\} +1. \label{eq:SAT MSE}
\end{align}
\end{subequations}
\subsubsection{Optimal Equalizer and Weight Updates}
For any TBS user $k$, the interference-plus noise term is given by \cite{10054084,5756489}
\begin{equation} \label{eq:TBS-IN term}
    \mathrm{T}_{k} = \sum\nolimits_{i=1}^K|\mathbf{h}_{tr,k}\mathbf{w}_{b,i}|^2 + \sum\nolimits_{l=1}^L|\mathbf{g}_{_{SAT \to k}}\mathbf{w}_{s,l}|^2 + \sigma_k^2.
\end{equation}
The optimal WMMSE equalizer and weight are respectively given by
\begin{subequations} 
\begin{align}
    u_{k}^{\star} &\leftarrow \frac{\mathbf{h}_{tr,k}\mathbf{w}_{b,k}}{\mathrm{T}_{k}},\label{eq:TBS equalizer}\\
    \omega_{k}^{\star} &\leftarrow \frac{1}{e_k}. \label{eq:TBS weight}
\end{align}
\end{subequations}
Similarly, for any SAT user $l$, the interference-plus-noise power is expressed as
\begin{equation}\label{eq:SAT-IN term}
    \mathrm{T}_{l} = \sum\nolimits_{j=1}^L|\mathbf{h}_{s,l}\mathbf{w}_{s,j}|^2 + \sum\nolimits_{k=1}^K|\mathbf{g}_{_{TBS \to l}}\mathbf{w}_{b,k}|^2 + \sigma_l^2.
\end{equation}
and the corresponding equalizer and weight updates are
\begin{subequations}
\begin{align}
    u_{l}^{\star} &\leftarrow \frac{\mathbf{h}_{s,l}\mathbf{w}_{s,l}}{\mathrm{T}_{l}}, \label{eq:SAT equalizer}\\
    \omega_{l}^{\star} &\leftarrow \frac{1}{e_l}.\label{eq:SAT weight}
\end{align}
\end{subequations}
\subsubsection{Precoder Updates}
With the $\{u_k,u_l\}$ and $\{\omega_k,\omega_l\}$ fixed, the TBS precoder admit a closed-form solution given by \cite{10054084,5756489}
\begin{equation}\label{eq:TBS Precoder update}
    \mathbf{w}_{b,k}^{\star} = (\mathbf{A_b} + \lambda_b \mathbf{I}_{M_b})^{-1}\mathbf{b}_{b,k},
\end{equation}
where $\lambda_b \geq 0$ is the Lagrangian multiplier variable chosen via bisection to satisfy the power constraint
 \begin{equation}
     \sum\nolimits_{k=1}^K\|\mathbf{w}_{b,k}\|^2 = P_b.
 \end{equation}
 The matrix $\mathbf{A_b}$ and the vector $\mathbf{b}_{b,k}$ are defined as
\begin{align}
    \mathbf{A_b}\! =\!\! \sum\nolimits_{i=1}^K \!\!\!\omega_{i}|u_{i}|^2\mathbf{h}_{tr,i}^H\mathbf{h}_{tr,i} \!+\!\! \sum\nolimits_{j=1}^L\!\!\!\omega_j|u_j|^2\mathbf{g}_{_{TBS \to l}}^H\mathbf{g}_{_{TBS \to l}},\label{eq:Ab}
\end{align}
\begin{equation}\label{eq:b_k}
     \mathbf{b}_{b,k} = \omega_{k}u_{k}^*\mathbf{h}_{tr,k}^H \in \mathbb{C}^{M_b}.
\end{equation}
Similarly, the SAT precoders are updated as
\begin{equation}\label{eq:SAT Precoder update}
    \mathbf{w}_{s,l}^{\star} = (\mathbf{A_s} + \lambda_s \mathbf{I}_{M_s})^{-1}\mathbf{b}_{s,l},
\end{equation}
where $\lambda_s$ selected such that
 \begin{equation}
     \sum\nolimits_{l=1}^L\|\mathbf{w}_{s,l}\|^2 = P_s.
 \end{equation}
 Here,
\begin{align}
    \mathbf{A_s}\! =\!\! \sum\nolimits_{j=1}^L\!\! \omega_{j}|u_{j}|^2\mathbf{h}_{s,j}^H\mathbf{h}_{s,j} \!+\!\!\! \sum\nolimits_{i=1}^K\!\!\omega_i|u_i|^2\mathbf{g}_{_{SAT \to k}}^H\mathbf{g}_{_{SAT \to k}},\label{eq:As}
\end{align}
\begin{equation}\label{eq:b_l}
     \mathbf{b}_{s,l} = \omega_{l}u_{l}^*\mathbf{h}_{s,l}^H \in \mathbb{C}^{M_s}.
\end{equation}
\begin{algorithm}[t]

\caption{WMMSE-based transmit Precoder Optimization}
\label{alg:WMMSE} \small
 \textbf{Input:} $P_b$, $P_s$, $\boldsymbol{\Theta}_U$, $\boldsymbol{\Theta}_H$, $\mathbf{q}_{_U}$, $\mathbf{q}_{_H}$, $\epsilon_{\mathbf{w}}$, $t_{max}^{\mathbf{w}}$. \\
\textbf{Initialize:} Precoders $\mathbf{w}_{b,k}^{(0)},\mathbf{w}_{s,l}^{(0)}$, compute $\mathcal{R}^{(0)}$ and set $t\leftarrow0$
\begin{algorithmic}[1]
\REPEAT   
\STATE Compute $\{\mathrm{T}_k^{(t)}$,$\mathrm{T}_l^{(t)}\}$ as in \eqref{eq:TBS-IN term} and \eqref{eq:SAT-IN term} and update equalizers $u_k^{(t+1)}$ and $u_l^{(t+1)}$ using \eqref{eq:TBS equalizer} and \eqref{eq:SAT equalizer}.
\STATE Compute MSEs $\{e_k^{(t+1)},e_l^{(t+1)}\}$ as in \eqref{eq:TBS MSE} and \eqref{eq:SAT MSE} and update weights $\{\omega_k^{(t+1)},\omega_l^{(t+1)}\}$ using \eqref{eq:TBS weight} and \eqref{eq:SAT weight}.
\STATE Construct $\mathbf{A}_b^{(t+1)}$ and $\mathbf{b}_{b,k}^{(t+1)}$ using \eqref{eq:Ab} and \eqref{eq:b_k}. 
\STATE Construct $\mathbf{A}_s^{(t+1)}$ and $\mathbf{b}_{s,l}^{(t+1)}$ using \eqref{eq:As} and \eqref{eq:b_l}.
\STATE Find Lagrange multiplier variables $\{\lambda_b,\lambda_s\}$ using bisection method. 
\STATE Update TBS precoders
\[ \mathbf{w}_{b,k}^{(t+1)} = (\mathbf{A}_b^{(t+1)} + \lambda_b \mathbf{I})^{-1} \mathbf{b}_{b,k}^{(t+1)}. \]
\STATE Update SAT precoders
\[ \mathbf{w}_{s,l}^{(t+1)} = (\mathbf{A}_s^{(t+1)} + \lambda_s \mathbf{I})^{-1} \mathbf{b}_{s,l}^{(t+1)}. \]
\STATE $t \leftarrow t+1$
\UNTIL $\left|\mathcal{R}^{(t)}-\mathcal{R}^{(t-1)}\right|\le \epsilon_{\mathbf{w}}$ or $t = t_{max}^{\mathbf{w}}$
\end{algorithmic}
\textbf{Output:} Optimized precoders $\{\mathbf{w}_{b,k}^\star, \mathbf{w}_{s,l}^\star\}$.
\end{algorithm}
The WMMSE algorithm for precoder optimization is summarized in \textbf{Algorithm~1}, where $\epsilon_{\mathbf{w}}$ and $t_{max}^{\mathbf{w}}$ denotes the convergence tolerance and the  maximum number of iterations, respectively. According to \cite{5756489}, \textbf{Algorithm~1} is guaranteed to converge to a stationary point of subproblem~{\textbf{(P3)}}.\vspace{-1.25em}
\subsection{UAV-RIS and HAP-RIS Phase Shift Optimization}
For fixed UAV/HAP trajectories and optimized precoders obtained from the previous subproblem, the corresponding BCD subproblem reduces to the optimization of UAV-RIS and HAP-RIS phase shift matrices. These subproblems are further reduced to unit-modulus constrained quadratic minimization problems, which are solved on the complex oblique (unit-modulus) manifold using the RCG method. 
Let $\mathbf{v}_{_U} \in \mathbb{C}^{N_U}$ and $\mathbf{v}_{_H} \in \mathbb{C}^{N_H}$ denote the UAV-RIS and HAP-RIS phase vectors with entries $\mathbf{v}_{_{U,i}} = e^{\iota\theta_{U,i}}$ and $\mathbf{v}_{_{H,j}} = e^{\iota\theta_{H,j}}$, respectively. 
\subsubsection{UAV-RIS phase shift optimization}
For any TBS user $k$, define
\begin{equation}\label{UAV-RIS_QF_Identities_TBS}
    \mathbf{a}_{k,i} = \operatorname{diag}(\mathbf{h}^*_{U,k})\mathbf{H}_{TBS,U}\mathbf{w}_{b,i}\,, \quad c_{k,i} = \mathbf{h}_{TBS,k}\mathbf{w}_{b,i}\,.
\end{equation}
Similarly, for any SAT user $l$, the TBS-to-user $l$ interference components that depend on UAV-RIS can be defined as
\begin{equation}\label{UAV-RIS_QF_Identities_SAT}
    \mathbf{\tilde{a}}_{l,i} = \operatorname{diag}(\mathbf{h}^*_{U,l})\mathbf{H}_{TBS,U}\mathbf{w}_{b,i}\,, \quad \tilde{c}_{l,i} = \mathbf{h}_{TBS,l}\mathbf{w}_{b,i}\,.
\end{equation}
From the WMMSE formulation, the MSE for the TBS user $k$ is
\begin{align}
    J_k^{(tr)} =  \omega_k\Big[|u_k|^2\Big(&\!\sum\nolimits_{i=1}^K|\mathbf{h}_{tr,k}\mathbf{w}_{b,i}|^2  \!+\!\! \sum\nolimits_{l=1}^L|\mathbf{g_{_{SAT \to k}}}\mathbf{w}_{s,l}|^2\nonumber\\&\hspace{2em} + \sigma_k^2\Big)-2\Re\{u_k\mathbf{h}_{tr,k}\mathbf{w}_{b,k}\}+1\Big].\label{Weighted_MSE_k_UAV}
\end{align}
Substituting (\ref{UAV-RIS_QF_Identities_TBS}) into (\ref{Weighted_MSE_k_UAV}) and expanding yields a quadratic form in $\mathbf{v}_{_U}$
\begin{equation}\label{Weighted_MSE_k_QF}
    J_k^{(tr)} = \mathbf{v}_{_U}^H\mathbf{Q}_{U,k}^{(tr)}\mathbf{v}_{_U} - 2\Re\left\{(\mathbf{\mathfrak{q}}_{_{U,k}}^{(tr)})^H\mathbf{v}_{_U}\right\} + \text{const}_k^{(tr)},
\end{equation}
where 
\begin{subequations}
\begin{align}
    \mathbf{Q}_{U,k}^{(tr)} &= \omega_k|u_k|^2\sum\nolimits_{i=1}^K\mathbf{a}_{k,i}\mathbf{a}_{k,i}^H, \\
    \mathbf{\mathfrak{q}}_{_{U,k}}^{(tr)} &= \omega_k\Big(u_k^*\mathbf{a}_{k,k}-|u_k|^2\sum\nolimits_{i=1}^Kc^*_{k,i}\mathbf{a}_{k,i}\Big),    
\end{align}
\end{subequations}
\begin{align}
    \text{const}_k^{(tr)} = \omega_k\Big[|u_k|^2\Big(&\sum\nolimits_{i=1}^K|c_{k,i}|^2 + \!\!\sum\nolimits_{l=1}^L|\mathbf{g_{_{SAT \to k}}}\mathbf{w}_{s,l}|^2\nonumber\\& + \sigma_k^2\Big) -2\Re\{u_kc_{k,k}\}+1\Big].
\end{align}
Although the desired SAT-to-$l$ channel is independent of the UAV-RIS, the interference of the TBS to any SAT user $l$ depends on it. Consequently, the MSE for the SAT user $l$ is
\begin{align}
    J_l^{(s,in)} =  \omega_l\Big[|u_l|^2\Big(\!&\sum\nolimits_{j=1}^L|\mathbf{h}_{s,l}\mathbf{w}_{s,j}|^2 \!+\!\!\sum\nolimits_{i=1}^K|\mathbf{g}_{_{TBS \to l}}\mathbf{w}_{b,i}|^2 \nonumber\\&+ \sigma_l^2\Big)-2\Re\{u_l\mathbf{h}_{s,l}\mathbf{w}_{s,l}\}+1\Big].\label{Weighted_MSE_l_UAV}
\end{align}
Substituting (\ref{UAV-RIS_QF_Identities_SAT}) into (\ref{Weighted_MSE_l_UAV}) gives
\begin{equation}\label{Weighted_MSE_l_QF_SAT}
    J_l^{(s,in)} = \mathbf{v}_{_U}^H\mathbf{Q}_{U,l}^{(s)}\mathbf{v}_{_U} - 2\Re\left\{(\mathbf{\mathfrak{q}}_{_{U,l}}^{(s)})^H\mathbf{v}_{_U}\right\} + \text{const}_l^{(s)},
\end{equation}
where
\begin{subequations}
\begin{align}
    \mathbf{Q}_{U,l}^{(s)} = \omega_l|u_l|^2\sum\nolimits_{i=1}^K\tilde{\mathbf{a}}_{l,i}\tilde{\mathbf{a}}_{l,i}^H, \\
    \mathbf{\mathfrak{q}}_{_{U,l}}^{(s)} = \omega_l|u_l|^2\sum\nolimits_{i=1}^K\tilde{c}^*_{l,i}\tilde{\mathbf{a}}_{l,i},    
\end{align}
\end{subequations}
and
\begin{align}
 \text{const}_l^{(s)} = \omega_l\Big[|u_l|^2\Big(&\sum\nolimits_{i=1}^K|\tilde{c}_{l,i}|^2 + \sum\nolimits_{j=1}^L|\mathbf{h}_{s,l}\mathbf{w}_{s,j}|^2 \nonumber\\&+ \sigma_l^2\Big) -2\Re\{u_l\mathbf{h}_{s,l}\mathbf{w}_{s,l}\}+1\Big].
\end{align}
By collecting the contributions of all users, the UAV-RIS phase shift subproblem can be expressed as 
\begin{subequations}\label{UAV_Problem_Final}
    \begin{align}
      \textbf{(P4)}\quad  \min_{\mathbf{v}_{_U}} &\quad f(\mathbf{v}_{_U})=\mathbf{v}_{_U}^H\mathbf{Q}_U\mathbf{v}_{_U}-2\Re\{\mathbf{\mathfrak{q}}^H_{_U}\mathbf{v}_{_U}\}\\
    \text{s.t.} &\quad|\mathbf{v}_{_{U,i}}| =1, \quad\forall \,i,
    \end{align}
\end{subequations}
where
\begin{subequations}
    \begin{align}
        \mathbf{Q}_U &= \sum\nolimits_{k=1}^K\mathbf{Q}_{U,k}^{(tr)}+\sum\nolimits_{l=1}^L\mathbf{Q}_{U,l}^{(s)},\\
        \mathbf{\mathfrak{q}}_{_U} &= \sum\nolimits_{k=1}^K\mathbf{\mathfrak{q}}_{_{U,k}}^{(tr)}-\sum\nolimits_{l=1}^L\mathbf{\mathfrak{q}}_{_{U,l}}^{(s)}.
    \end{align}
\end{subequations}
\subsubsection{HAP-RIS phase shift optimization}
For any SAT user $l$, define
\begin{equation}\label{HAP-RIS_QF_Identities}
    \mathbf{a}_{l,j} = \operatorname{diag}(\mathbf{h}^*_{H,l})\mathbf{H}_{SAT,H}\mathbf{w}_{s,j}, \quad c_{l,j} = \mathbf{h}_{SAT,l}\mathbf{w}_{s,j}.
\end{equation}
Similarly, for any TBS user $k$, the SAT-to-$k$ interference components that depend on the HAP-RIS can be defined as
\begin{equation}\label{HAP-RIS_QF_Identities_interf}
    \mathbf{\tilde{a}}_{k,j} = \operatorname{diag}(\mathbf{h}^*_{H,k})\mathbf{H}_{SAT,H}\mathbf{w}_{s,j}, \;\; \tilde{c}_{k,j} = \mathbf{h}_{SAT,k}\mathbf{w}_{s,j}.
\end{equation}
From the WMMSE formulation, the MSE for SAT user $l$ is
\begin{align}
    J_l^{(s)} =  \omega_l\Big[|u_l|^2\Big(&\sum\nolimits_{j=1}^L|\mathbf{h}_{s,l}\mathbf{w}_{s,j}|^2  \!+\!\! \sum\nolimits_{k=1}^K|\mathbf{g_{_{TBS \to l}}}\mathbf{w}_{b,k}|^2 \nonumber\\&+ \sigma_l^2\Big) -2\Re\{u_l\mathbf{h}_{s,l}\mathbf{w}_{s,l}\}+1\Big].\label{Weighted_MSE_l_HAP}
\end{align}
Substituting (\ref{HAP-RIS_QF_Identities}) into (\ref{Weighted_MSE_l_HAP}) and expanding yields a quadratic form in $\mathbf{v}_{_H}$
\begin{equation}\label{Weighted_MSE_l_QF_HAP}
    J_l^{(s)} = \mathbf{v}_{_H}^H\mathbf{Q}_{H,l}^{(s)}\mathbf{v}_{_H} - 2\Re\left\{(\mathbf{\mathfrak{q}}_{_{H,l}}^{(s)})^H\mathbf{v}_{_H}\right\} + \text{const}_{_{H,l}}^{(s)},
\end{equation}
where 
\begin{subequations}
\begin{align}
    \mathbf{Q}_{H,l}^{(s)} = \omega_l|u_l|^2\sum\nolimits_{j=1}^L\mathbf{a}_{l,j}\mathbf{a}_{l,j}^H, \\
    \mathbf{\mathfrak{q}}_{H,l}^{(s)} = \omega_l\Big(u_l^*\mathbf{a}_{l,l}-|u_l|^2\sum\nolimits_{j=1}^Lc^*_{l,j}\mathbf{a}_{l,j}\Big),    
\end{align}
\end{subequations}
\begin{align}
        \text{const}_{_{H,l}}^{(s)} = \omega_l\Big[|u_l|^2\Big(\sum\nolimits_{j=1}^L&|c_{l,j}|^2 + \sum\nolimits_{k=1}^K|\mathbf{g_{_{TBS \to l}}}\mathbf{w}_{b,k}|^2 \nonumber\\&+ \sigma_l^2\Big) -2\Re\{u_lc_{l,l}\}+1\Big].
\end{align}
Although the desired TBS-to-user $k$ channel is independent of the HAP-RIS, the interference from the SAT to the TBS user $k$ depends on it. Consequently, the MSE for the TBS user $k$ is
\begin{align}
    J_k^{(tr,in)} =  \omega_k\Big[|u_k|^2\Big(&\sum\nolimits_{i=1}^K\!|\mathbf{h}_{tr,k}\mathbf{w}_{b,i}|^2\nonumber  +\!\! \sum\nolimits_{j=1}^L\!|\mathbf{g}_{_{SAT \to k}}\mathbf{w}_{s,j}|^2\\&+\!\! \sigma_k^2\Big)\!\!-\!2\Re\{u_k\mathbf{h}_{tr,k}\mathbf{w}_{b,k}\}\!+\!\!1\!\Big],\label{Weighted_MSE_k_HAP}
\end{align}
Substituting (\ref{HAP-RIS_QF_Identities_interf}) into (\ref{Weighted_MSE_k_HAP}) gives
\begin{equation}\label{Weighted_MSE_l_QF}
    J_k^{(tr,in)} = \mathbf{v}_{_H}^H\mathbf{Q}_{H,k}^{(str)}\mathbf{v}_{_H} - 2\Re\left\{(\mathbf{\mathfrak{q}}_{_{H,k}}^{(tr)})^H\mathbf{v}_{_H}\right\} + \text{const}_{_{H,k}}^{(tr)},
\end{equation}
where
\begin{subequations}
\begin{align}
    \mathbf{Q}_{H,k}^{(tr)} = \omega_k|u_k|^2\sum\nolimits_{j=1}^L\tilde{\mathbf{a}}_{k,j}\tilde{\mathbf{a}}_{k,j}^H, \\
    \mathbf{\mathfrak{q}}_{H,k}^{(tr)} = \omega_k|u_k|^2\sum\nolimits_{j=1}^L\tilde{c}^*_{k,j}\tilde{\mathbf{a}}_{k,j},    
\end{align}
\end{subequations}
\begin{align}
        \text{const}_{_{H,k}}^{(tr)} = \omega_k\Big[|u_k|^2&\Big(\sum\nolimits_{j=1}^L|\tilde{c}_{k,j}|^2 + \sum\nolimits_{i=1}^K|\mathbf{h}_{tr,k}\mathbf{w}_{b,i}|^2 \nonumber\\&+ \sigma_k^2\Big) -2\Re\{u_k\mathbf{h}_{tr,k}\mathbf{w}_{b,k}\}+1\Big].
\end{align}
By collecting the contributions of all user, the HAP-RIS phase shift subproblem can be expressed as
\begin{subequations}\label{HAP_Problem_Final}
    \begin{align}
        \textbf{(P5)}\quad\min_{\mathbf{v}_{_H}} &\quad f(\mathbf{v}_{_H})=\mathbf{v}_{_H}^H\mathbf{Q}_H\mathbf{v}_{_H}-2\Re\{\mathbf{\mathfrak{q}}^H_{_H}\mathbf{v}_{_H}\}\\
    \text{s.t.} &\quad|\mathbf{v}_{_{H,j}}| =1, \quad\forall j,
    \end{align}
\end{subequations}
where
\begin{subequations}
    \begin{align}
        \mathbf{Q}_H &= \sum\nolimits_{l=1}^L\mathbf{Q}_{H,l}^{(s)}+\sum\nolimits_{k=1}^K\mathbf{Q}_{H,k}^{(tr)},\\
        \mathbf{\mathfrak{q}}_{_H} &= \sum\nolimits_{l=1}^L\mathbf{\mathfrak{q}}_{_{H,l}}^{(s)}-\sum\nolimits_{k=1}^K\mathbf{\mathfrak{q}}_{_{H,k}}^{(tr)}.
    \end{align}
\end{subequations}
The subproblems \textbf{(P4)} and \textbf{(P5)} share an identical mathematical structure and can be expressed in the following unified form
\begin{subequations}\label{Unified_RIS_Problem}
\begin{align}
\textbf{(P6)}\quad\min_{\mathbf{v}\in\mathbb{C}^{\mathsf{N}}}
&\quad f(\mathbf{v}) = \mathbf{v}^H\mathbf{Q}\mathbf{v}
-2\Re\{\mathbf{\mathfrak{q}}^H\mathbf{v}\} \\
\text{s.t.}
&\quad |\mathbf{v}_\mathfrak{n}| = 1,\quad \mathfrak{n}=1,\dots,\mathsf{N},
\end{align}
\end{subequations}
where $\mathbf{v}$ denotes the RIS phase shift vector and $\mathsf{N}$ denotes the number of passive RIS elements, with
\[
(\mathbf{v},\mathbf{Q},\mathbf{\mathfrak{q}},\mathsf{N}) =
\begin{cases}
(\mathbf{v}_{_U},\mathbf{Q}_U,\mathbf{\mathfrak{q}}_{_U},N_U), & \text{UAV-RIS},\\
(\mathbf{v}_{_H},\mathbf{Q}_H,\mathbf{\mathfrak{q}}_{_H},N_H), & \text{HAP-RIS}.
\end{cases}
\]
The unit-modulus constraints in \eqref{Unified_RIS_Problem} define a complex oblique manifold
\begin{equation}\label{Manifold}
\mathcal{M} \triangleq
\left\{
\mathbf{v}\in\mathbb{C}^{\mathsf{N}} :
|\mathbf{v}_\mathfrak{n}|=1,\ \forall \,\mathfrak{n}
\right\}.
\end{equation}
The tangent space at $\mathbf{v}^{(t)}\in\mathcal{M}$ is given by
\begin{equation}\label{Tangent}
\mathcal{T}_{\mathbf{v}^{(t)}}\mathcal{M}
=
\left\{
\boldsymbol{\eta}\in\mathbb{C}^{\mathsf{N}} :
\Re\{\boldsymbol{\eta}\circ(\mathbf{v}^{(t)})^*\}=0
\right\}.
\end{equation}
The Euclidean gradient of $f(\mathbf{v})$ is
\begin{equation}\label{Euclidean space}
\nabla f(\mathbf{v}) = 2\mathbf{Q}\mathbf{v} - 2\mathbf{\mathfrak{q}},
\end{equation}
and the corresponding Riemannian gradient is obtained via orthogonal projection as
\begin{equation}\label{Riemmanian gradient}
\mathrm{grad}_{\mathbf{v}^{(t)}} f =
\nabla f(\mathbf{v}^{(t)})
-
\Re\!\left\{
\nabla f(\mathbf{v}^{(t)}) \circ (\mathbf{v}^{(t)})^*
\right\}
\circ \mathbf{v}^{(t)}.
\end{equation}
The Riemannian conjugate gradient update is given by
\begin{equation}\label{Unified_RCG_Update}
\boldsymbol{\zeta}^{(t+1)}=
-\mathrm{grad}_{\mathbf{v}^{(t)}} f
+
\chi^{(t+1)}
\mathsf{T}_{\mathbf{v}^{(t)}\rightarrow \mathbf{v}^{(t+1)}}(\boldsymbol{\zeta}^{(t)}),
\end{equation}
where $\chi^{(t+1)}$ is the Polak-Ribière parameter \cite[Ch. ~8.3]{AbsilMahonySepulchre+2008}, and
$\mathsf{T}_{\mathbf{v}^{(t)}\rightarrow \mathbf{v}^{(t+1)}}(\cdot)$ denotes the vector transport, defined as
\begin{equation}\label{Transport}
\mathsf{T}_{\mathbf{v}^{(t)}\to \mathbf{v}^{(t+1)}}(\boldsymbol{\zeta})
=
\boldsymbol{\zeta}
-
\Re\{\boldsymbol{\zeta}\circ(\mathbf{v}^{(t+1)})^*\}
\circ \mathbf{v}^{(t+1)}.
\end{equation}
A retraction is applied to ensure feasibility
\begin{equation}\label{Retraction}
\mathsf{R}_{\mathbf{v}^{(t)}}(\xi^{(t)}\boldsymbol{\zeta}^{(t)})
=
\mathrm{unt}\big(\mathbf{v}^{(t)}+\xi^{(t)}\boldsymbol{\zeta}^{(t)}\big),
\end{equation}
where $\mathrm{unt}(\cdot)$ denotes element-wise normalization and $\xi^{(t)}$ is obtained via Armijo backtracking line search \cite[Ch. ~4.2]{AbsilMahonySepulchre+2008}.

\begin{algorithm}[t]
\caption{ Manifold-based RCG for UAV-RIS and HAP-RIS Phase-shift Optimization }
\label{alg:RCG} \small
\textbf{Input:} $\mathcal{M}$, $f(\mathbf{v})$, $\epsilon_{_\mathcal{M}}$, $t_{\max}^{\mathcal{M}}$.

\textbf{Initialize:} t $\leftarrow$ 0; randomly select $\mathbf{v}^{(t)}$, $\xi^{(t)}$, $\chi^{(t)}$; $\zeta^{(t)} = -\mathrm{grad}_{\mathbf{v}^{(t)}} f(\mathbf{v})$; 

\begin{algorithmic}[1]
\REPEAT
    \STATE Update $\mathbf{v}^{(t+1)}$ using retraction with line search based step size.
    \STATE Calculate the Riemannian gradient at $\mathbf{v}^{(t+1)}$, Polak-Ribière parameter, and transport from $\mathbf{v}^{(t)}$ to $\mathbf{v}^{(t+1)}$.
    \STATE Update the conjugate search direction using \eqref{Unified_RCG_Update}.
    \STATE $t \leftarrow t+1$
\UNTIL $\| \mathrm{grad}f(\mathbf{v})\|\leq \epsilon_{_\mathcal{M}}$ or $t = t_{\max}^{\mathcal{M}}$
\end{algorithmic}
\textbf{Output:} Optimized RIS phase vector $\mathbf{v}^\star$.
\end{algorithm}
The manifold-based optimization procedure for RIS phase shift design is summarized in \textbf{Algorithm~2}, where $\epsilon_{\mathcal{M}}$ and $t_{max}^{\mathcal{M}}$ denotes a predefined convergence threshold and maximum number of iterations, respectively. According to~\cite{AbsilMahonySepulchre+2008}, \textbf{Algorithm~2} is guaranteed to converge to a critical point of the objective function \eqref{Unified_RIS_Problem}, at which the Riemannian gradient vanishes.\vspace{-1em}
\subsection{UAV and HAP Trajectory Optimization}
With the precoding vectors and RIS phase shift matrices obtained from the previously discussed subproblems, the UAV and HAP trajectory optimization subproblem can be decoupled from \textbf{(P1)} and formulated as
\begin{subequations} \label{Objective-Trajectory}
\begin{align}
\textbf{(P7)}\quad&\max_{\substack{ \{\mathbf{q}_{_U}[n],\mathbf{q}_{_H}[n]\}}} 
 \frac{1}{N}\sum_{n=1}^{N}\! \Big( \sum_{k=1}^{K} R_k[n] + \!\!\sum_{l=1}^{L} R_l[n] \Big) \\
&\text{s.t.} \;\;\; 
\eqref{opt:UAV trajectories},\eqref{opt:HAP trajectories},\eqref{opt:UAV altitude},\eqref{opt:HAP altitude},\eqref{opt:UAV init-final},\eqref{opt:HAP init-final}. 
\end{align}
\end{subequations}
This trajectory subproblem remains non-convex due to the distance-dependent channel terms in the objective function. To address this challenge, we develop an SCA framework that constructs convex affine surrogates of the received power expressions. 
As the positions of TBS, SAT, and user are fixed, only the cascaded TBS-to-UAV-RIS-to-user $k$ and SAT-to-HAP-RIS-to-user $l$ links depend on the UAV and HAP trajectories, respectively. The direct TBS-user $k$ and SAT-user $l$ links are independent of trajectories and are therefore excluded from trajectory optimization. 
\subsubsection{Surrogate Construction for the UAV-assisted Cascaded Link}
For any TBS user $k$, the TBS-to-UAV-RIS-to-user link involves two distances and two corresponding path-loss exponents, $d_1=d_{TBS,U}$ and $d_2 = d_{U,k}$,with $\gamma_1 =\alpha_{_{TBS,U}} $ and $\gamma_2=\alpha_{_{U,k}}$, respectively. The corresponding received signal power can be expressed as
\begin{equation}
    \mathbf{p}(\mathbf{q}_{_U}) = C_U d_1^{-\gamma_1}d_2^{-\gamma_2}, 
\end{equation}
where $C_U >0$ collects all distance-independent terms, including precoder gains, RIS phase shifts, array responses, Rician fading components, and transmit power, and remains fixed during each SCA iteration 
Let $d_1^{(t)},d_2^{(t)}$ denote the distances at iteration $t$. We define
\begin{equation}
    h(d_1,d_2) = d_1^{-\gamma_1}d_2^{-\gamma_2}.
\end{equation}
The partial derivatives of $h(d_1,d_2)$ are given by
\begin{equation}
    \begin{aligned}
        \frac{\partial h}{\partial d_1} = -\gamma_1d_1^{-\gamma_1-1}d_2^{-\gamma_2}, \quad
        \frac{\partial h}{\partial d_2} = -\gamma_2d_1^{-\gamma_1}d_2^{-\gamma_2-1}.
    \end{aligned}
\end{equation}
Applying the first-order Taylor expansion yields the affine lower bound as \cite{9656117}
\begin{equation}
    h(d_1,d_2) \geq h^{(t)} + h_{d_1}^{(t)}(d_1-d_1^{(t)}) + h_{d_2}^{(t)}(d_2-d_2^{(t)}),
\end{equation}
where $h^{(t)} = {d_1^{(t)}}^{-\gamma_1}{d_2^{(t)}}^{-\gamma_2}$, $h_{d_1}^{(t)} = -\gamma_1{d_1^{(t)}}^{-\gamma_1-1}{d_2^{(t)}}^{-\gamma_2}$, and $h_{d_2}^{(t)}=-\gamma_2{d_1^{(t)}}^{-\gamma_1}{d_2^{(t)}}^{-\gamma_2-1}$.

Next, substituting $d_1=\|\mathbf{q}_{_U}-\mathbf{q}_{_{TBS}}\|$ and  $d_2 = \|\mathbf{q}_{_U}-\mathbf{q}_k\|$, we linearize the Euclidean norm at $\mathbf{q}_{_U}^{(t)}$ as \cite{9656117}
\begin{subequations}
    \begin{align}
        \|\mathbf{q}_{_U}-\mathbf{q}_{_{TBS}}\| &\geq d_1^{(t)} + \frac{(\mathbf{q}_{_U}^{(t)}-\mathbf{q}_{_{TBS}})^T}{d_1^{(t)}}(\mathbf{q}_{_U}-\mathbf{q}_{_U}^{(t)}),\label{Eq:UAV Linear Norms1}\\
        \|\mathbf{q}_{_U}-\mathbf{q}_k\| &\geq d_2^{(t)} + \frac{(\mathbf{q}_{_U}^{(t)}-\mathbf{q}_{_k})^T}{d_2^{(t)}}(\mathbf{q}_{_U}-\mathbf{q}_{_U}^{(t)}).\label{Eq:UAV Linear Norms2}
    \end{align}
\end{subequations}
Finally, substituting \eqref{Eq:UAV Linear Norms1} and \eqref{Eq:UAV Linear Norms2} into the Taylor surrogate gives the affine approximation of the received power \cite{9656117}
    \begin{align}
        \hat{\mathbf{p}}^{(t)}(\mathbf{q}_{_U}) = C_U \bigg( &h^{(t)} + h_{d_1}^{(t)}\frac{(\mathbf{q}_{_U}^{(t)}-\mathbf{q}_{_{TBS}})^T}{d_1^{(t)}}(\mathbf{q}_{_U}-\mathbf{q}_{_U}^{(t)})\nonumber \\
         &+ h_{d_2}^{(t)}\frac{(\mathbf{q}_{_U}^{(t)}-\mathbf{q}_{_k})^T}{d_2^{(t)}}(\mathbf{q}_{_U}-\mathbf{q}_{_U}^{(t)})\bigg).\label{Eq:UAV Power Surrogate}
    \end{align}

\begin{algorithm}[t]\small{
\caption{SCA-based UAV and HAP Trajectory Optimization}
\label{alg:SCA} \small
\textbf{Input:} $\mathbf w_{b,k},\mathbf w_{s,l}$, $\mathbf v_{_U},\mathbf v_{_H}$, $V_{U,max},V_{H,max}$, $\epsilon_\mathbf{q}$, $t_{max}^\mathbf{q}$.\\
\textbf{Initialization:} Initial trajectories
$\mathbf q_{_U}^{(0)},\mathbf q_{_H}^{(0)}$, and set iteration index $t \leftarrow 0$
\begin{algorithmic}[1]
\REPEAT
\STATE Compute distances $d_{TBS,U}^{(t)}, d_{U,k}^{(t)}, \tilde d_{SAT,H}^{(t)}, \tilde d_{H,l}^{(t)}$
\STATE Construct affine power surrogates
$\hat{\mathbf p}^{(t)}(\mathbf q_{_U}[n])$ and
$\hat{\mathbf p}^{(t)}(\mathbf q_{_H}[n])$ using
(\ref{Eq:UAV Power Surrogate}) and (\ref{Eq:HAP Power Surrogate})

\STATE Form surrogate SINR and rate expressions
$\hat R_k[n], \hat R_l[n]$

\STATE Solve convex problem (\ref{Subproblem-trajectory})
to obtain $\mathbf q_{_U}^{(t+1)}, \mathbf q_{_H}^{(t+1)}$

\STATE $t \leftarrow t + 1$
\UNTIL {$\|\mathbf q_{_U}^{(t)}-\mathbf q_{_U}^{(t-1)}\|_2\le\epsilon_q$
\AND 
$\|\mathbf q_{_H}^{(t)}-\mathbf q_{_H}^{(t-1)}\|_2\le\epsilon_q$}\\
or $t=t_{max}^q$

\STATE $\mathbf q_{_U}^\star \leftarrow \mathbf q_{_U}^{(t)}, \quad
\mathbf q_{_H}^\star \leftarrow \mathbf q_{_H}^{(t)}$
\end{algorithmic}
\textbf{Output:} Optimized trajectories $\{\mathbf q_{_U}^\star,\mathbf q_{_H}^\star\}$.}
\end{algorithm}

\subsubsection{Surrogate Construction for the HAP-assisted Cascaded Link}
For the SAT-to-HAP-RIS-to-user $l$ link, let $\tilde{d}_1$ and $\tilde{d}_2$ denote the SAT-to-HAP-RIS distance, and the HAP-RIS-to-user $l$ distance, respectively. Upon repeating the same procedure as that for the UAV-assisted cascade link, the affine lower-bound surrogate of the received power is given by \cite{9656117}
\begin{align}
\hat{\mathbf{p}}^{(t)}(\mathbf{q}_{_H}) = C_H\bigg(& \tilde{h}^{(t)} 
+ \tilde{h}_{\tilde{d}_1}^{(t)}\frac{(\mathbf{q}_{_H}^{(t)}-\mathbf{q}_{_{SAT}})^T}{\tilde{d}_1^{(t)}}(\mathbf{q}_{_H}-\mathbf{q}_{_H}^{(t)})\nonumber\\
&+ \tilde{h}_{\tilde{d}_2}^{(t)}\frac{(\mathbf{q}_{_H}^{(t)}-\mathbf{q}_{_l})^T}{\tilde{d}_2^{(t)}}(\mathbf{q}_{_H}-\mathbf{q}_{_H}^{(t)})\bigg),\label{Eq:HAP Power Surrogate}
\end{align}
where $\tilde{h}^{(t)} = ({\tilde{d}_1^{(t)}})^{-2}({\tilde{d}_2^{(t)}})^{-2}, 
    \tilde{h}_{\tilde{d}_1}^{(t)} = -2({\tilde{d}_1^{(t)}})^{-3}({\tilde{d}_2^{(t)}})^{-2}, 
    \tilde{h}_{\tilde{d}_2}^{(t)}\\ =-2({\tilde{d}_2^{(t)}})^{-3}({\tilde{d}_1^{(t)}})^{-2}.$ The surrogates in (\ref{Eq:UAV Power Surrogate}) and (\ref{Eq:HAP Power Surrogate}) are affine in $\mathbf{q}_{_U}$ and $\mathbf{q}_{_H}$, respectively, and can therefore be directly incorporated into the SCA-based trajectory optimization framework.

\subsubsection{Surrogate SINR and Rate Expressions}
For each time slot $n$, we optimize the UAV and HAP trajectories $\big\{\mathbf{q}_{_U}[n],\mathbf{q}_{_H}[n]\big\}$. Let $\mathcal{I}_k[n]$ and $\mathcal{I}_l[n]$ denote the interference-plus-noise power at user $k$ and user $l$, respectively. 
\par Using affine power surrogates, the surrogate SINR of TBS user $k$ is
\begin{equation}
    \widehat{SINR}_k^{(t)}[n] = \frac{\hat{\mathbf{p}}^{(t)}(\mathbf{q}_{_U}[n])}{\mathcal{I}_k[n]}.
\end{equation}
Thus, the corresponding surrogate rate is
\begin{align}
    \hat{R}_k[n]\! =\! \log_2\!\!\left(\!1\!+\!\widehat{SINR}_k^{(t)}[n]\!\right)\!=\!\log_2\!\!\left(\!1\!+\!\frac{\hat{\mathbf{p}}^{(t)}(\mathbf{q}_{_U}[n])}{\mathcal{I}_k[n]}\!\right).
\end{align}
Similarly, for any user $l$, the surrogate rate is 
\begin{equation}
    \hat{R}_l[n] =\log_2\left(1+\frac{\hat{\mathbf{p}}^{(t)}(\mathbf{q}_{_H}[n])}{\mathcal{I}_l[n]}\right).
\end{equation}
Accordingly, the convex subproblem can be formulated as
\begin{subequations} \label{Subproblem-trajectory}
    \begin{align}
       \textbf{(P8)} &\max_{\substack{\{\mathbf{q}_{_U}[n],\mathbf{q}_{_H}[n]\}}} \frac{1}{N}\!\!\sum\nolimits_{n=1}^{N}\!\! \Bigg(\!\!\! \sum\nolimits_{k=1}^{K}  \!\!\hat{R}_k[n] \!+\!\!\! \sum\nolimits_{l=1}^{L} \! \!\hat{R}_l[n] \!\!\Bigg) \\
&\text{s.t.} \;\;\; 
\eqref{opt:UAV trajectories},\eqref{opt:HAP trajectories},\eqref{opt:UAV altitude},\eqref{opt:HAP altitude},\eqref{opt:UAV init-final},\eqref{opt:HAP init-final}.
    \end{align}
\end{subequations}
Finally, the subproblem \textbf{(P8)} is convex and can be efficiently solved using the CVX toolbox \cite{Boyd_Vandenberghe_2004}.
However, due to the successive convex approximations adopted in the above derivations, the optimal solution of \textbf{(P8)} is generally not equivalent to that of the original non-convex subproblem \textbf{(P7)}.
Therefore, \textbf{(P8)} must be solved iteratively within an SCA framework to progressively approach a locally optimal solution of \textbf{(P7)}.
\par The SCA-based optimization procedure for the UAV and HAP trajectories is summarized in \textbf{Algorithm~3}, where $\epsilon_\mathbf{q}$ and $t_{max}^\mathbf{q}$ denote the convergence tolerance and the maximum number of iterations, respectively.

\begin{algorithm}[t]\small{
\caption{BCD-based Joint Precoder, UAV-RIS \& HAP-RIS Phase-shift, and Trajectory Optimization}
\label{alg:BCD} \small
\textbf{Input:} System parameters, channel statistics, power budgets $\{P_b,P_s\}$, tolerance $\epsilon$,
maximum iterations $I_{\max}$ \\
\textbf{Initialization:} 
$\{\mathbf w_{b,k}^{(0)},\mathbf w_{s,l}^{(0)}\}$,
$\{\mathbf v_U^{(0)},\mathbf v_H^{(0)}\}$,
$\{\mathbf q_{_U}^{(0)},\mathbf q_{_H}^{(0)}\}$,
set $i \leftarrow 0$
\begin{algorithmic}[1]
\WHILE{$i \le I_{\max}$}
\STATE Update effective channels based on $\mathbf w_{b,k}^{(i)},\mathbf w_{s,l}^{(i)},\mathbf v_{_U}^{(i)},\mathbf v_{_H}^{(i)},\mathbf q_{_U}^{(i)},\mathbf q_{_H}^{(i)}$
\STATE \textbf{Precoder Optimization:} By using \textbf{Algorithm}~\ref{alg:WMMSE} to obtain
$\mathbf w_{b,k}^{(i+1)},\mathbf w_{s,l}^{(i+1)}$
\STATE \textbf{RIS Phase Optimization:} By using \textbf{Algorithm}~\ref{alg:RCG} to obtain
$\mathbf v_{_U}^{(i+1)},\mathbf v_{_H}^{(i+1)}$
\STATE \textbf{Trajectory Optimization:} By using \textbf{Algorithm}~\ref{alg:SCA} to obtain
$\mathbf q_{_U}^{(i+1)},\mathbf q_{_H}^{(i+1)}$


\STATE Compute the average sum-rate $\mathcal R^{(i+1)}$
\IF{$\big|\mathcal R^{(i+1)} - \mathcal R^{(i)}\big| \le \epsilon$}
\STATE \textbf{break}
\ENDIF

\STATE $i \leftarrow i+1$
\ENDWHILE
\end{algorithmic}
\textbf{Output:} Optimized precoders $\{\mathbf w_{b,k}^\star,\mathbf w_{s,l}^\star\}$,
RIS phase vectors $\{\mathbf v_{_U}^\star,\mathbf v_{_H}^\star\}$,
and UAV/HAP trajectories $\{\mathbf q_{_U}^\star,\mathbf q_{_H}^\star\}$.}
\end{algorithm}

\subsection{Overall Algorithm}
To jointly optimize the transmit precoders, RIS phase shifts, and UAV/HAP trajectories, we employ a BCD framework \cite{11030800}, in which the original non-convex problem is decomposed into three tractable subproblems that are alternatively optimized. Specifically, for fixed UAV-RIS and HAP-RIS phase shifts and trajectories, the transmit precoder optimization subproblem is solved using the WMMSE algorithm, which guarantees a monotonic improvement of the average sum-rate and converges to a stationary point.
For fixed UAV/HAP trajectories and precoders, the UAV-RIS and HAP-RIS phase shift optimization subproblem is solved using manifold-based RCG method, which efficiently handles the unit-modulus constraints of the RIS elements.
For fixed precoders and RIS phase shift matrices, the UAV and HAP trajectory optimization subproblem is solved using SCA, where a sequence of convex surrogates are solved iteratively. 
At the $i$-th BCD iteration, the effective channels are first updated according to the current RIS phase shifts and trajectories. Subsequently, the transmit precoders, RIS phase vectors, and trajectories are updated in sequence using the WMMSE, manifold-based RCG, and SCA algorithms, respectively. Each subproblem is solved to a prescribed accuracy, controlled by individual convergence tolerances and maximum iteration numbers. The BCD iterations terminate when the increment of the average sum-rate between two consecutive iterations falls below a predefined threshold or when the maximum number of BCD iterations is reached. The overall procedure is summarized in \textbf{Algorithm~4}.

\textit{Note:} Due to space limitations, the convergence behavior and computational complexity analysis of the proposed algorithm are omitted.


\begin{table*}[t]
\centering
\caption{Simulation Parameters}
\label{tab:sim_param}
\begin{tabular}{|p{4.6cm}|p{2.1cm}|p{4cm}|p{3.5cm}|}
\hline
\textbf{Parameter} & \textbf{Value} & \textbf{Parameter} & \textbf{Value} \\ \hline
TBS antennas  & $M_b=8$ & SAT antennas  & $M_s=32$ \\
TBS users  & $K=3$ & SAT users  & $L=4$ \\
UAV-RIS elements  & $N_U=16$ & HAP-RIS elements  & $N_H=36$ \\
UAV altitude range & $z_{_U}=[80,250]$ m & HAP altitude range & $z_{_H}=[17,25]$ km  \\
Rician factor (TBS-to-user $k$) & $\kappa_{_{TBS,k}}=5$ & Rician factor (SAT-to-user $l$) & $\kappa_{_{SAT,l}}=3$\\
Rician factor (TBS-to-UAV-RIS) & $\kappa_{_{TBS,U}}=8$ & Rician factor (SAT-to-HAP-RIS) & $\kappa_{_{SAT,H}}=6$\\
Rician factor (UAV-RIS-to-user $k$) & $\kappa_{_{U,k}}=5$ & Rician factor (HAP-RIS-to-user $l$) & $\kappa_{_{H,l}}=3$\\
TBS carrier frequency  & $f_c=3.5$ GHz & SAT carrier frequency  & $\check{f}_c=20$ GHz \\
TBS transmit power  & $P_b=44.77$ dBm & SAT transmit power  & $P_s=54.8$ dBm\\
Max. UAV speed  & $V_{U,max}\!=\!30$ m/s & Max. HAP speed & $V_{H,max}=5$ m/s \\
Time slots  & $N=60$ & Slot duration  & $\delta=1$ s \\
SAT altitude  & $H_{SAT}=600$ km & Noise power & $\sigma_k^2\!\!=\!\!\sigma_l^2\!\!=\!\!\sigma^2\!=\!-90$ dBm \cite{10929715}  \\
Reference path-loss  & $\beta_o=0$ dBm \cite{9656117} & SAT antenna gain  & $G_{SAT}=30$ dBi \cite{3GPP_TR_38_821} \\
Path-loss exponent (TBS-to-user $k$) & $\alpha_{_{TBS,k}}=3.5$ & HAP antenna gain  & $G_{HAP}=14$ dBi \cite{3GPP_TR_38_821}\\
Path-loss exponent (TBS-to-UAV-RIS) & $\alpha_{_{TBS,U}}=2.2$ & User antenna gain  & $G_l=0$ dBi\\
Path-loss exponent (UAV-RIS-to-user $k$) & $\alpha_{_{U,k}}=2.2$ & Rain attenuation mean \& variance & ($\mu_r$,$\sigma_r^2$) \!=\! ($-2.6,1.63$) dB \cite{10768935}\\ \hline
\end{tabular}
\vspace{-1.25em}
\end{table*}

 \begin{figure*}[t!]
\centering
\begin{subfigure}{0.335\textwidth} 
\centerline{\includegraphics[scale=0.43] {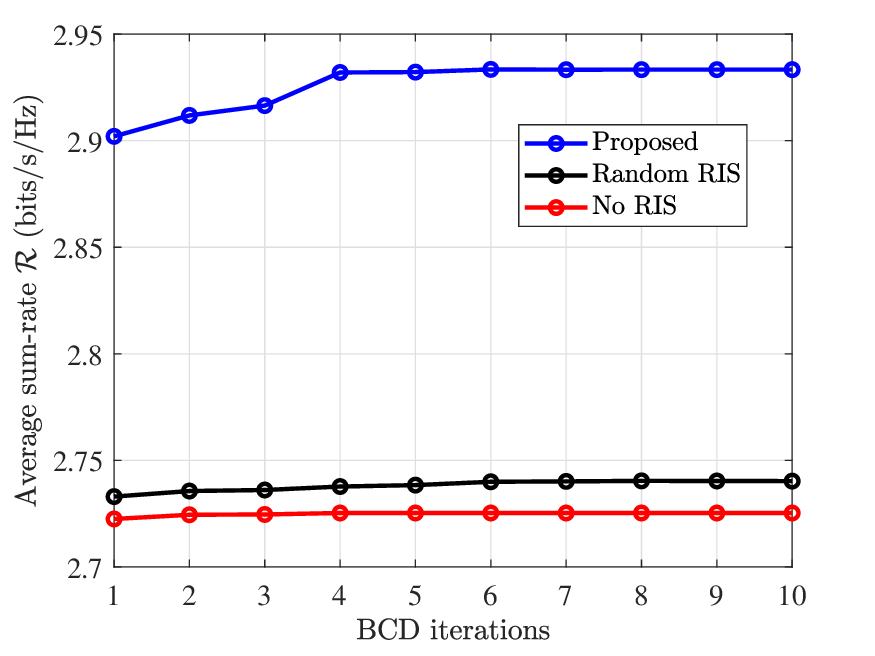}}
\caption{Convergence of proposed \textbf{Algorithm~4}.}
    \label{fig:BCDConv}
\end{subfigure}%
\begin{subfigure}{0.335\textwidth}
\centerline{\includegraphics[scale=0.43]{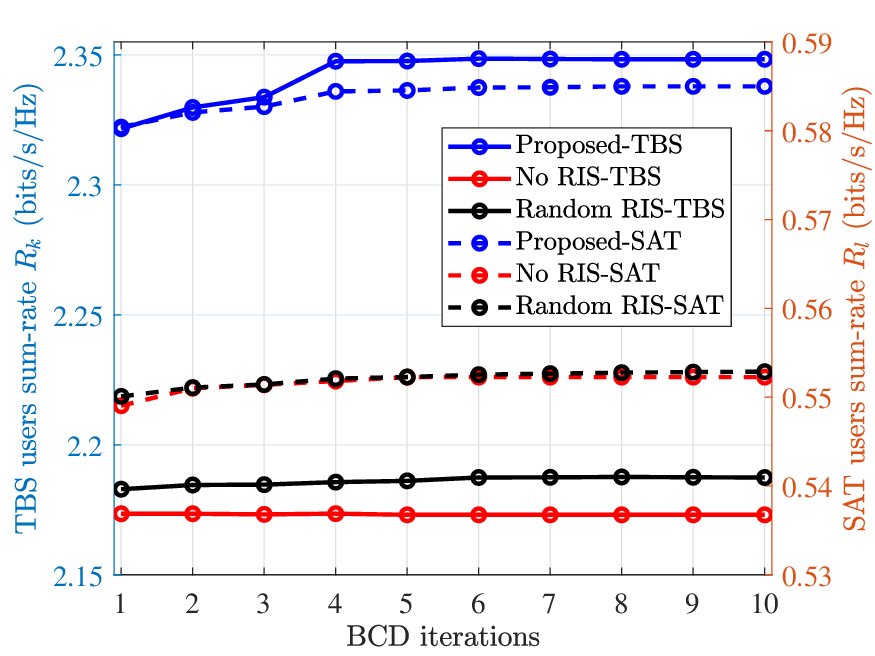}}
\caption{Convergence for $R_k$ and $R_l$.}
    \label{fig:BCDConv_TBS_SAT}
\end{subfigure}%
\begin{subfigure}{0.335\textwidth}
\centerline{\includegraphics[scale=0.43]{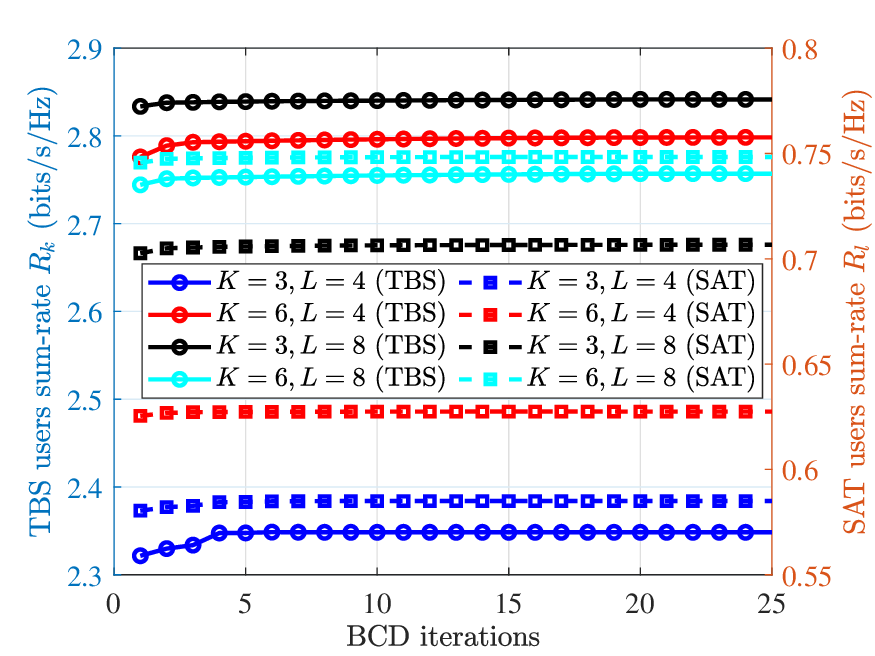}}
\caption{Convergence for different $K$ and $L$.}
    \label{fig:BCDConv_TBS_SAT_diffUsers}
\end{subfigure}
\caption{Convergence of the proposed \textbf{Algorithm}~\ref{alg:BCD}, convergence of the TBS and SAT user sum-rates, $(R_k)$ \& $(R_l)$, and convergence behaviour under different number of TBS users $K$ and SAT users $L$.}
\label{fig:combined1}
\vspace{-1.25em}
\end{figure*}

\section{Simulation Results}\label{ref:Simulation results}
In this section, we present simulation results to evaluate the performance of the proposed BCD-based optimization framework and compare it with benchmark schemes. 
The fixed positions of TBS and SAT are set as $\mathbf{q}_{_{TBS}} = (0,0,0)$ m, and $\mathbf{q}_{_{SAT}} = (0.6,0.8,600)$ km, respectively. 
The TBS users are independently and uniformly distributed within a rectangular area in the ground plane, with horizontal coordinates $x_{_k} \in [150,400]$ m and $y_{_k}\in[130,400]$ m. In parallel, SAT users are uniformly distributed within a separate ground region defined by $x_{_l}\in[1100,1400]$ m and $y_{_l}\in[600,900]$ m. All users are assumed to be located at zero altitude. The UAV and HAP are initialized with straight-line trajectories at fixed altitudes of 165 m and 21 km, respectively. The UAV starts its trajectory in the vicinity of the TBS toward the TBS user region, while the HAP performs a short horizontal displacement around the centroid of SAT users. Both trajectories are uniformly discretized over $N$ time slots. The simulation parameters are presented in Table \ref{tab:sim_param}. We considered the following benchmarks: no RIS, where the RIS is removed from the system, random RIS, where the RIS is deployed but its phase shifts are fixed and randomly generated, without optimization, and Fixed trajectory, where the UAV moves along a predetermined straight-line path with a constant speed and without trajectory optimization.

\par Fig.~\ref{fig:BCDConv} illustrates the convergence behavior of the proposed BCD-based optimization framework in terms of the average sum-rate versus the number of BCD iterations.
The proposed algorithm exhibits fast and stable convergence, reaching a locally optimal solution in approximately six iterations, demonstrating the effectiveness of the BCD framework in jointly optimizing transmit precoders, RIS phase shifts, and UAV/HAP trajectories. In contrast, the benchmark schemes show nearly unchanged sum-rate performance across iterations, which reflects their limited capability to exploit the controllable propagation environment. As a result, the proposed joint design achieves an approximately $7.05\%$ higher average sum-rate compared to the random RIS case. The observed performance gain is attributed to the coherent phase alignment enabled by the optimized RIS configuration, which enhances the effective channel gain and improves the received SINR. The monotonic increase of the objective value across iterations indicates that each subproblem contributes positively to the overall optimization objective, ensuring reliable convergence of the proposed algorithm. 
\par Fig.~\ref{fig:BCDConv_TBS_SAT} depicts the convergence of the average sum-rate for the TBS and SAT links versus the number of BCD iterations. It can be observed that the proposed algorithm converges rapidly for both links, demonstrating the numerical stability and effectiveness of the adopted BCD framework. In contrast, the random RIS and no RIS benchmark schemes achieve significantly lower sum-rates with nearly constant performance across iterations, indicating their limited ability to exploit channel characteristics. Specifically, the proposed joint design achieves an approximately $7.36\%$ and $5.8 \%$ higher average sum-rate compared to the random RIS case for TBS and SAT, respectively.
The performance advantage of the proposed method arises from the joint optimization of the transmit precoders, RIS phase shifts, and UAV/HAP trajectories, which enables constructive signal combining and improves the effective channel gains for both terrestrial and satellite links. The monotonic improvement in the objective values further validates that each subproblem is well designed and consistently enhances the global objective. It can also be observed that the performance gap between the proposed joint design and the benchmark methods remains consistent after convergence, highlighting the robustness of the proposed framework in simultaneously improving the terrestrial and satellite sum-rate performance.

 \begin{figure*}[t!]
\centering
\begin{subfigure}{0.335\textwidth} 
\centerline{\includegraphics[scale=0.43] {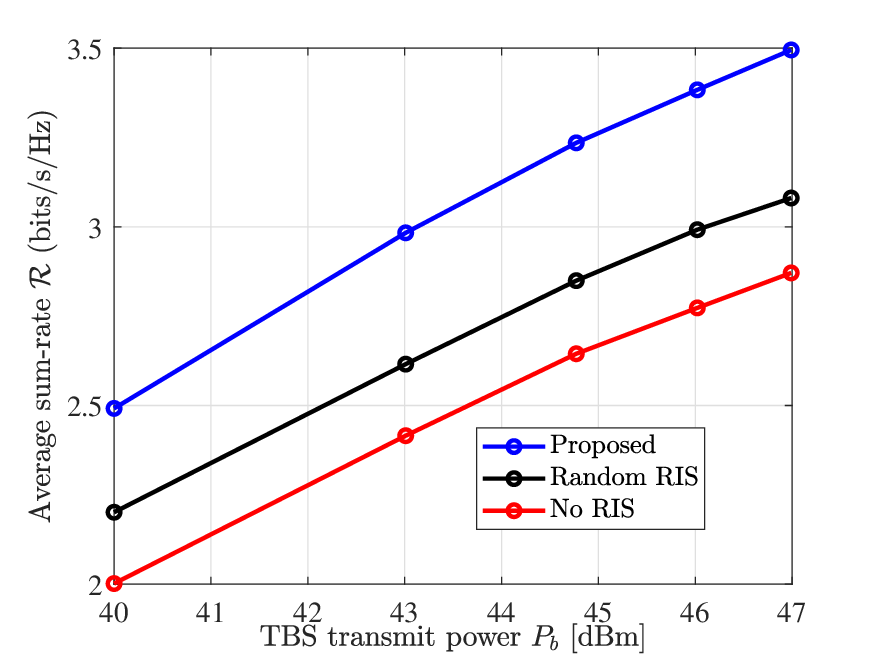}}
    \caption{Average sum-rate versus $P_b$ [dBm].}
    \label{fig:Sum-rate versus TBS Transmit power}
\end{subfigure}%
\begin{subfigure}{0.335\textwidth}
\centerline{\includegraphics[scale=0.43]{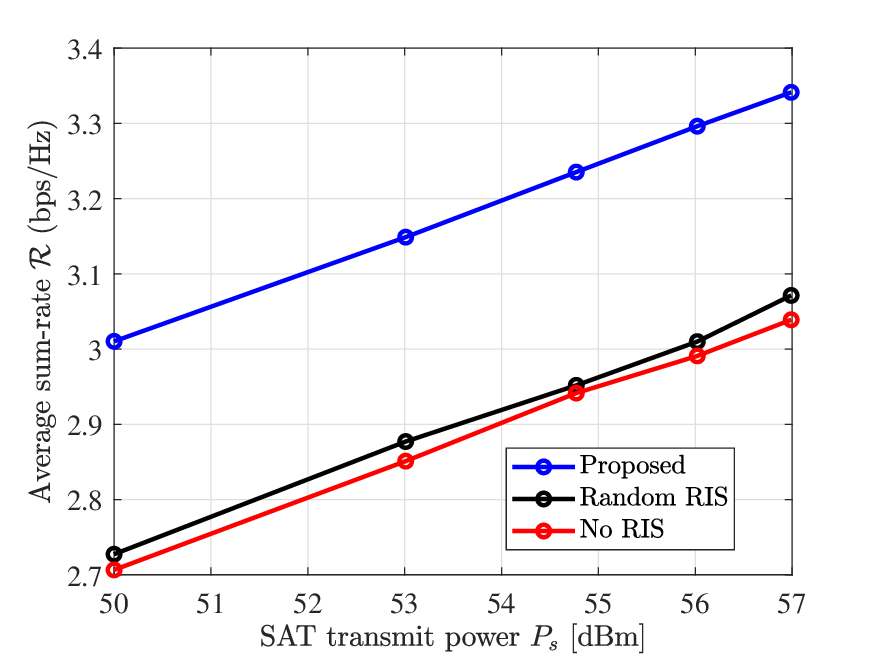}}
    \caption{Average sum-rate versus $P_s$ [dBm].}
    \label{fig:Sum-rate versus SAT Transmit power}
\end{subfigure}%
\begin{subfigure}{0.335\textwidth}
\centerline{\includegraphics[scale=0.43]{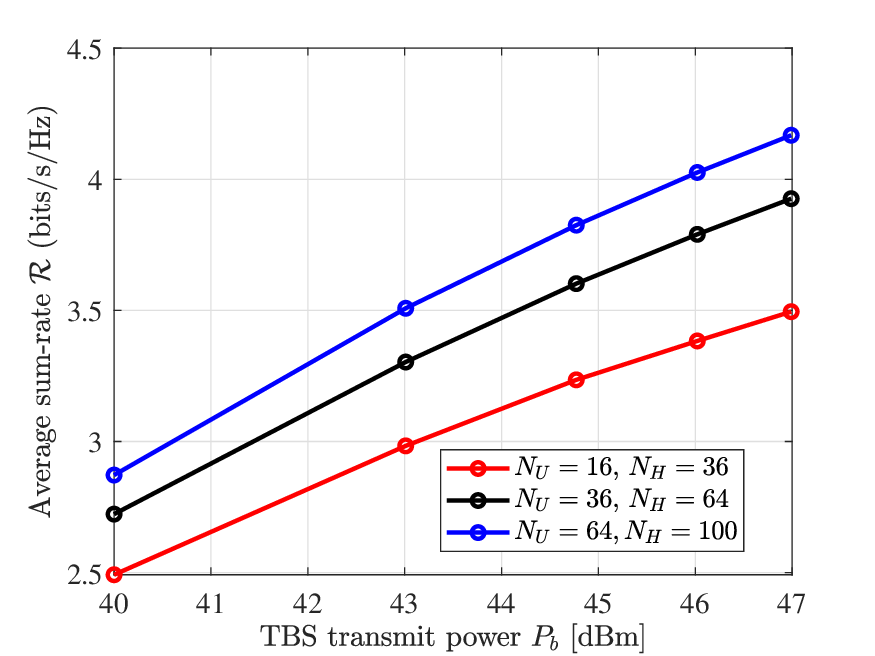}}
    \caption{$\mathcal{R}$ versus $P_b$ [dBm] for different $N_U$ and $N_H$.}
    \label{fig:different RIS}
\end{subfigure}
\caption{Impact of $P_b$ [dBm] and $P_s$ [dBm] on the average sum-rate, and average sum-rate versus $P_b$ for different numbers of UAV-RIS and HAP-RIS elements.}

\label{fig:combined2}
\vspace{-1.25em}
\end{figure*}

\par Fig.~\ref{fig:BCDConv_TBS_SAT_diffUsers} presents the convergence behavior under different configurations of the TBS and SAT user, $(K,L) \in \{(3,4),(6,4),(3,8),(6,8)\}$. The proposed algorithm converges within a few iterations for all configurations, demonstrating stable convergence performance irrespective of network scale. The sum-rate achieved increases with larger values of $K$ and $L$, as serving more users and satellite links provides additional multiuser spatial multiplexing gain and enhances spectral efficiency. As the number of users increases from $(K,L)=(3,4)$, the average sum-rate improves by about $17.4\%$ and $27.9\%$ for TBS and SAT, respectively. This result verifies that the proposed BCD-based framework is scalable and robust with respect to the system dimensions and can efficiently accommodate increased user density and satellite load without compromising convergence speed or stability.




\par Fig.~\ref{fig:Sum-rate versus TBS Transmit power} shows the average sum-rate versus the TBS transmit power ($P_b$) for the proposed joint design, the random RIS, and the no RIS benchmark. It can be observed that the average sum-rate increases monotonically with $P_b$ for all schemes, as the higher transmit power enhances the received signal strength at the users. The proposed joint design consistently outperforms the benchmark methods across the entire considered power range, while the random RIS scheme achieves moderate gains compared to the no RIS case.
The performance improvement of the proposed joint design results mainly from the optimized RIS phase shifts and the joint TBS-SAT transmission design, which significantly enhances the effective channel gain. In particular, as $P_b$ increases, the sum-rate improvement does not saturate rapidly, indicating that the system does not become strictly interference-limited in the high-SNR regime. This behavior suggests that the proposed design effectively exploits additional transmit power by improving signal alignment and mitigating multi-user interference. Consequently, the performance gap between the proposed joint design and the benchmark methods remains noticeable at high $P_b$. Specifically, when $P_b$ increases from $40$ to $47$ dBm, the proposed joint design achieves an average sum-rate improvement of approximately $40.25\%$, while $P_s$ is fixed at $54.77$ dBm.

\par Fig.~\ref{fig:Sum-rate versus SAT Transmit power} illustrates the average sum-rate versus SAT transmit power ($P_s$) for the proposed joint design, the random RIS, and the no RIS benchmark. It can be observed that the average sum-rate increases monotonically with $P_s$ for all schemes, as the higher satellite transmit power enhances the received signal strength at the users.
The proposed joint design consistently attains the highest sum-rate, whereas the random RIS scheme outperforms the no RIS benchmark by exploiting partial control over the propagation environment.
In the high-$P_s$ regime, the sustained performance gain of the proposed joint design indicates that the system remains power-limited rather than interference-limited.
This is attributed to the joint design and coordinated TBS–SAT transmission
As a result, the proposed joint design continues to benefit from increasing satellite transmit power, and the performance gap with respect to the benchmark schemes.
In particular, as $P_s$ increases from $50$ to $57$ dBm, the proposed joint design achieves an average sum-rate improvement of approximately $11\%$, while $P_b$ is fixed at $44.77$ dBm.

\par Fig.~\ref{fig:different RIS} illustrates the average sum-rate as a function of the TBS transmit power $P_b$ for different numbers of UAV-RIS and HAP-RIS elements, denoted by $(N_U$, $N_H)$. It is observed that the average sum rate increases monotonically with $P_b$ for all considered configurations, which reflects the improvement in SINR as the available transmit power increases. In addition, configurations with larger values of $N_U$ and $N_H$ consistently achieve higher sum-rates across the entire power range. For example, for $N_U = 64$ and $N_H=100$, the average sum-rate increases from $2.872$ bits/s/Hz at $P_b = 40$ dBm to $4.168$ bits/s/Hz at $P_b = 47$ dBm, corresponding to an improvement of approximately $45.13\%$. This behavior arises from the increased precoder and reflection gains provided by a larger number of RIS elements, which enhance the effective channel quality. The results indicate that jointly increasing the number of UAV-RIS and HAP-RIS elements significantly improves the system performance.



\begin{figure*}[t!]
\centering
\begin{subfigure}{0.335\textwidth}
\centerline{\includegraphics[scale=0.43]{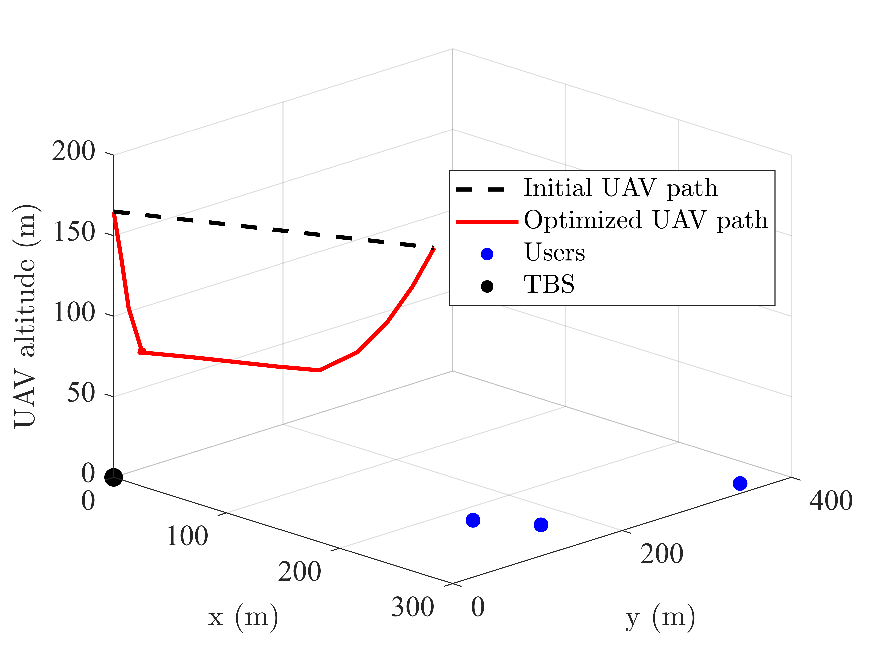}}
    \caption{Optimal UAV trajectory.}
    \label{fig:Optimal UAV trajectory}
\end{subfigure}%
\begin{subfigure}{0.335\textwidth}
\centerline{\includegraphics[scale=0.43]{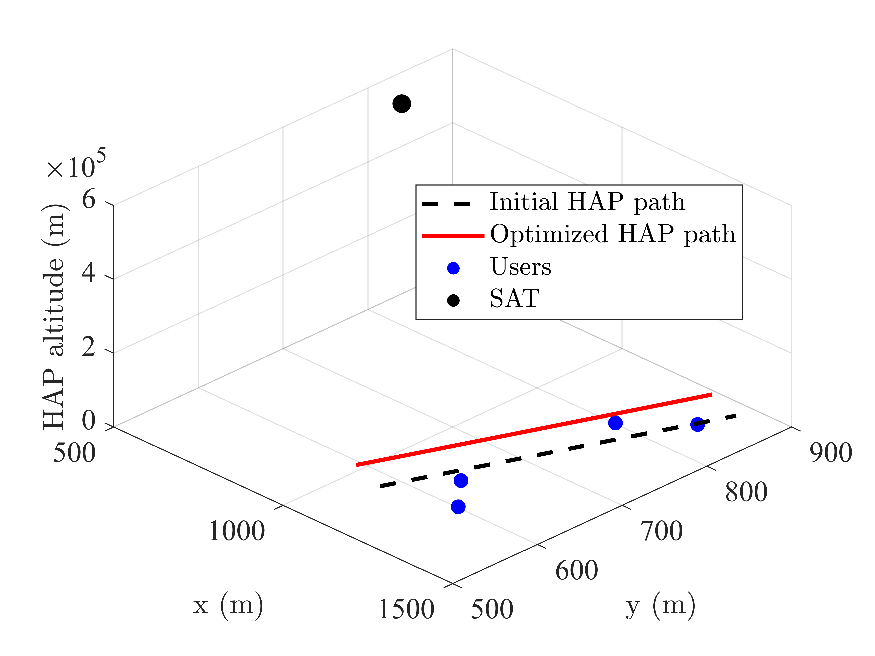}}
    \caption{Optimal HAP trajectory.}
    \label{fig:Optimal HAP trajectory}
\end{subfigure}%
\begin{subfigure}{0.335\textwidth} 
\centerline{\includegraphics[scale=0.43]{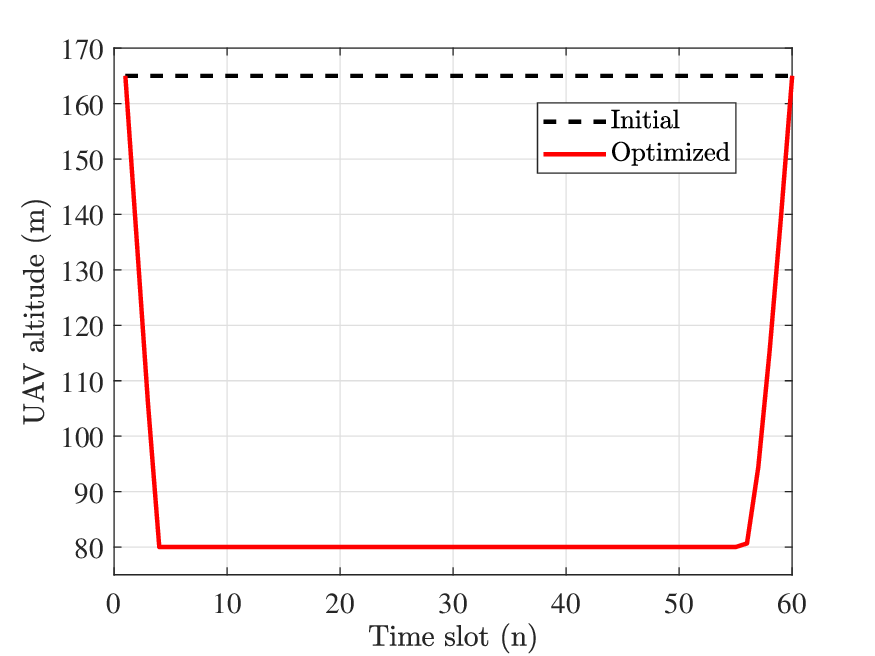}}
    \caption{UAV altitude versus $n$.}
    \label{fig:UAV_altitude}
\end{subfigure}
\caption{Optimal UAV and HAP trajectories, and impact of time slot $n$ on UAV altitude.}
\label{fig:combined3}
\vspace{-1.25em}
\end{figure*}

\begin{figure*}[t!]
\centering
\begin{subfigure}{0.335\textwidth}
\centerline{\includegraphics[scale=0.43]{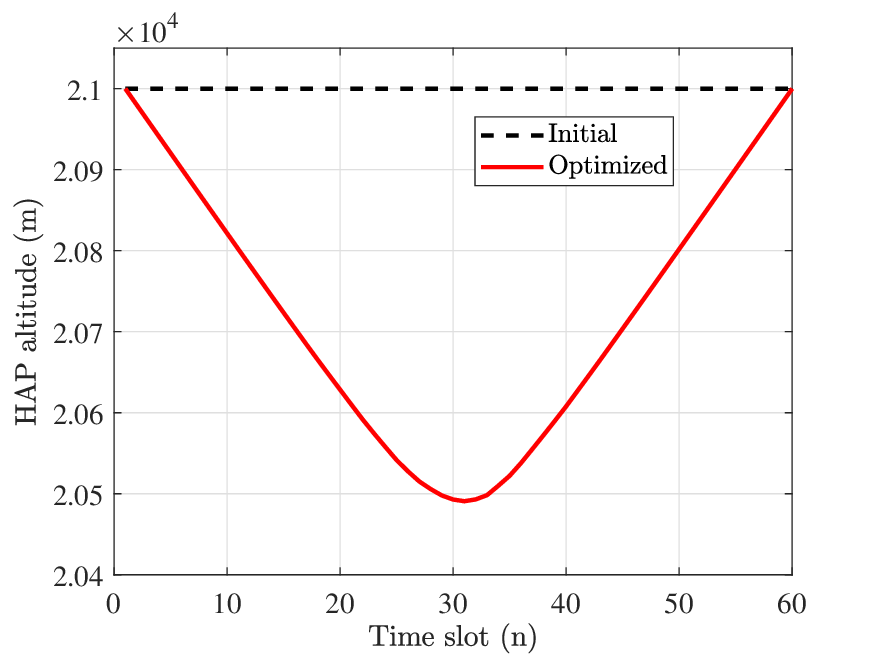}}
    \caption{HAP altitude versus $n$.}
    \label{fig:HAP_altitude}
\end{subfigure}%
\begin{subfigure}{0.335\textwidth} 
\centerline{\includegraphics[scale=0.43]{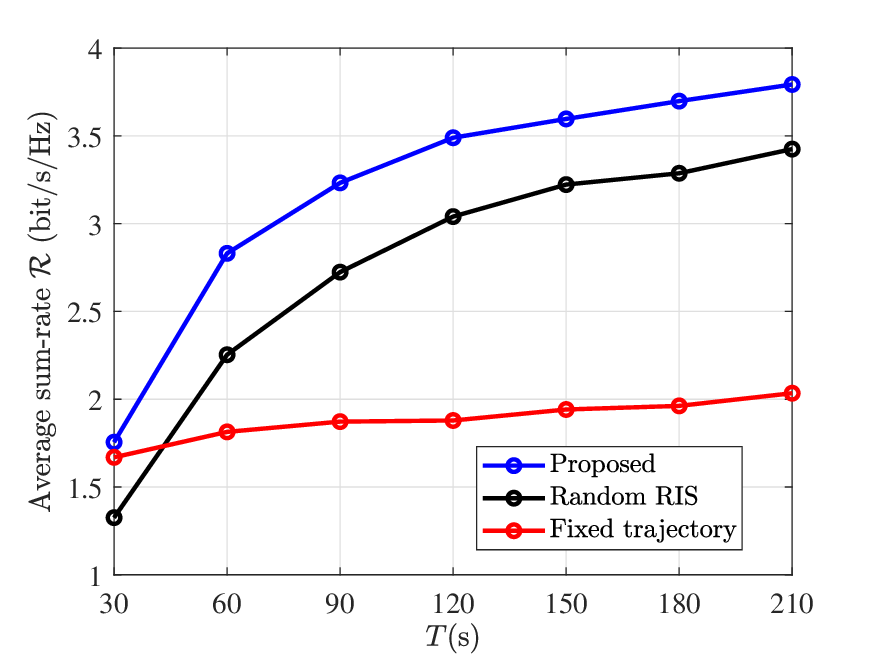}}
    \caption{Average sum-rate versus $T(s)$.}
    \label{fig:SRvsT}
\end{subfigure}%
\begin{subfigure}{0.335\textwidth}
\centerline{\includegraphics[scale=0.43]{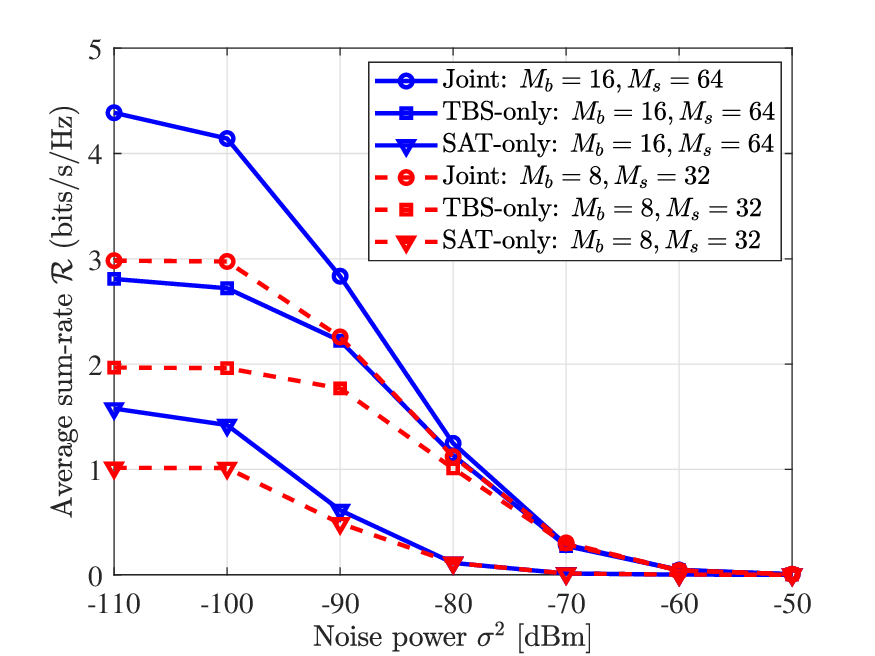}}
    \caption{Average sum-rate versus $\sigma^2$ [dBm].}
    \label{fig:SRvsNP}
\end{subfigure}
\caption{Impact of time slot $(n)$ on HAP altitude, and impact of flight period $T(s)$ and noise power $(\sigma^2)$ [dBm] on average sum-rate.}
\label{fig:combined4}
\vspace{-1.25em}
\end{figure*}

\par Fig.~\ref{fig:Optimal UAV trajectory} illustrates the optimal 3D UAV trajectory obtained by the proposed joint design, together with the initial UAV path, the positions of the users, and the TBS position. It can be observed that the optimized UAV trajectory significantly deviates from the initial path, allowing the UAV to adapt its horizontal position and altitude to better serve the ground users. The UAV moves closer to densely located users and adjusts its altitude to balance coverage and channel quality, thereby improving the effective communication links.
The optimized trajectory reflects the joint consideration of communication performance and mobility constraints within the proposed framework. By dynamically adjusting the UAV position in response to user positions, the system achieves improved channel conditions and enhanced sum-rate performance compared to the initial fixed straight-line trajectory.

\par Fig.~\ref{fig:Optimal HAP trajectory} illustrates the optimal 3D HAP trajectory obtained by the proposed joint design, together with the initial HAP path, the user positions, and the satellite position. It can be observed that the optimized HAP trajectory slightly displaced from the initial straight-line path and adapts its horizontal movement to better align with the spatial distribution of the users. By adjusting its position in response to user positions, the HAP achieves more favorable propagation conditions and improved coverage. This result demonstrates the effectiveness of jointly optimizing the HAP trajectory within the proposed joint design.

\par Fig.~\ref{fig:UAV_altitude} shows the UAV altitude variation versus the time slot index (n). The dashed curve represents the initial UAV path altitude, without trajectory optimization, while the solid curve corresponds to the optimized altitude obtained from the proposed joint design. It can be observed that the optimized UAV trajectory operates at a lower altitude during most of the time slots and increases its altitude near the beginning and the end of the mission. This altitude adaptation allows the UAV to improve the air-to-ground link quality with the users while respecting mobility and altitude constraints. The result highlights the benefit of incorporating UAV altitude optimization in enhancing the overall sum-rate performance.

\par Fig.~\ref{fig:HAP_altitude} shows the HAP altitude variation versus the time slot index (n). The dashed curve corresponds to the initial HAP path, while the solid curve represents the optimized altitude obtained using the proposed framework. It can be observed that the optimized HAP altitude follows a smooth convex profile, decreasing during the intermediate time slots and increasing toward the beginning and the end of the mission. This altitude adjustment enables the HAP to balance coverage and path-loss effects over time, resulting in more favorable propagation conditions. The result highlights the advantage of incorporating HAP altitude optimization in improving overall system performance.

\par In Fig.~\ref{fig:SRvsT}, the average sum-rate of the proposed joint design is compared with the random RIS and fixed-trajectory benchmark schemes for different flight periods $T$. It can be observed that the average sum-rate increases monotonically with $T$ for both the proposed and the random RIS schemes and gradually saturates. This behavior is attributed to the fact that a longer flight period allows the UAV and HAP to remain closer to the users for a longer duration, thereby exploiting more favorable channel conditions and improving service quality. In contrast, the fixed-trajectory scheme exhibits only marginal performance improvement, as the UAV and HAP move uniformly along a predetermined straight path, which limits their ability to adapt to channel variations.
The superiority of the proposed joint design is clearly demonstrated by the noticeable performance gap with respect to the benchmark schemes. Specifically, the proposed joint design achieves an approximately $10.75 \%$ higher average sum-rate compared to the random RIS case. A substantial performance gain is also observed relative to the fixed-trajectory scheme. These results show that both intelligent RIS configuration and trajectory optimization play a crucial role in enhancing the overall sum-rate performance of the system.

\par Fig.~\ref{fig:SRvsNP} illustrates the average sum-rate versus noise power $(\sigma^2)$ for different transmission strategies and antenna configurations. It can be observed that the average sum-rate decreases monotonically with increasing noise power for all considered schemes, reflecting the degradation of the received SINR under stronger noise conditions. The joint transmission strategy consistently outperforms the TBS-only and SAT-only schemes across the entire noise power range, demonstrating the benefit of simultaneously exploiting terrestrial and satellite links. In addition, larger antenna configurations $(M_b=16, M_s=64)$ achieve higher sum-rates than smaller ones $(M_b=8, M_s=32)$, highlighting the role of increased spatial degrees of freedom in mitigating noise effects.
The performance gain of the joint design can be attributed to the combined diversity and array gains offered by coordinated TBS–SAT transmission, which improves the robustness of the system against noise impairments. As the noise power increases beyond a certain level, the sum-rates of all schemes approach zero, indicating a noise-limited regime in which the benefits of joint transmission and larger antenna arrays become marginal under severe noise conditions.

\section{Conclusion}\label{ref:Conclusion}
This paper investigated a dual-aerial RIS-assisted ITNTN, where a UAV-RIS and an HAP-RIS jointly support terrestrial and satellite communications. An average sum-rate maximization problem was formulated by jointly optimizing the transmit precoders at the TBS and SAT, RIS phase shift, and the 3D trajectories of the UAV and HAP under practical constraints. To address the resulting non-convex problem, a BCD-based framework combining WMMSE, RCG, and SCA techniques was developed, and its convergence to a stationary point was established. The simulation results demonstrated that the proposed joint design significantly outperforms the benchmark schemes and achieves an approximately $7.05\%$ higher average sum-rate compared to the random RIS scheme, highlighting the importance of jointly optimizing communication and mobility resources in ITNTNs. Future research directions include extending the proposed framework to scenarios with imperfect channel state information, multiple UAVs or HAPs, and additional performance objectives such as energy efficiency, latency, and robustness.
\bibliographystyle{IEEEtran}   
\bibliography{ref}     
\end{document}